\documentclass[superscriptaddress,twocolumn,amssymb,nofootinbib,10pt]{revtex4-2}
\usepackage[normalem]{ulem}
\usepackage[utf8]{inputenc}
\usepackage{comment}
\usepackage{graphicx}
\usepackage{graphics}
\usepackage{amsmath, amssymb}
\usepackage{graphics,bm}
\usepackage{graphicx}
\usepackage{bbold}
\usepackage{slashed}
\usepackage{feynmf}
\usepackage{wrapfig}
\usepackage{hyperref}
\usepackage{cancel}
\usepackage[usenames]{color}  

\usepackage[compat=1.1.0]{tikz-feynman}
\usepackage{tikz}
\usetikzlibrary{arrows,shapes}
\usetikzlibrary{trees}
\usetikzlibrary{matrix,arrows} 			
\usetikzlibrary{positioning}				
\usetikzlibrary{calc,through}				
\usepackage{pgffor}							

\newcommand{\be}{\begin{equation}}
\newcommand{\ee}{\end{equation}}
\newcommand{\beq}{\begin{equation}}
\newcommand{\eeq}{\end{equation}}

\newcommand{\bqa}{\begin{eqnarray}}
\newcommand{\eqa}{\end{eqnarray}}

\def\square{\vcenter{\vbox{\hrule height.4pt
          \hbox{\vrule width.4pt height4pt
          \kern4pt\vrule width.3pt}\hrule height.4pt}}}

\usepackage[normalem]{ulem}

\begin{document}

\title{Chiral perturbation theory and Bose-Einstein condensation in QCD}


\author{Jens O. Andersen}
\email{jens.andersen@ntnu.no}
\affiliation{Department of Physics, Faculty of Natural Sciences,NTNU, 
Norwegian University of Science and Technology, H{\o}gskoleringen 5,
N-7491 Trondheim, Norway}

\affiliation{Niels Bohr International Academy, Blegdamsvej 17, DK-2100 Copenhagen, Denmark}
\author{Martin Kj{\o}llesdal Johnsrud}

\email{martin.johnsrud@ds.mpg.de} 

\affiliation{Department of Physics, Faculty of Natural Sciences,NTNU, 
Norwegian University of Science and Technology, H{\o}gskoleringen 5,
N-7491 Trondheim, Norway}

\affiliation{Department of Living Matter Physics, Max Planck Institute of Dynamics and Self-Organization, Am Fa{\ss}berg 17, DE-37077 G{\"o}ttingen, Germany}

\author{Qing Yu}
\email{yuq@swust.edu.cn}
\affiliation{School of Mathematics and Physics, Southwest University of Science and Technology, Mianyang 621010, China}

\author{Hua Zhou}
\email{zhouhua@swust.edu.cn}
\affiliation{School of Mathematics and Physics, Southwest University of Science and Technology, Mianyang 621010, China}

\date{\today}

\begin{abstract}
We present recent results in three-flavor chiral perturbation theory at finite isospin $\mu_I$ 
and strangeness $\mu_{S}$ chemical potentials at zero temperature.
The tree-level spectrum for the mesons and gauge bosons in the pion-condensed phase is derived. 
The phase diagram to ${\cal O}(p^2)$ in the $\mu_I$--$\mu_S$ plane is mapped out with and without
electromagnetic effects. The phase diagram consists of a vacuum phase and three Bose-condensed phases with condensates of $\pi^{\pm}$, $K^{\pm}$, and $K^{0}/\bar{K}^0$, respectively.
Including electromagnetic interactions, the charged Bose-condensed phases become Higgs phases via the Higgs mechanism.  We calculate the pressure, energy density, isospin density, and speed of sound in the pion-condensed phase to ${\cal O}(p^4)$. The results are compared with recent lattice simulations and the agreement is very good for isospin chemical potentials up to approximately 180 MeV. Moreover, by integrating out the $s$-quark, we show that the thermodynamic quantities
can be mapped onto their two-flavor counterparts with renormalized parameters.
The breaking of the $\mathrm{U}(1)$ symmetry in the Bose-condensed phases
gives rise to a Goldstone boson, whose dispersion is linear for small momenta. We use Son's prescription to construct an effective theory for the Goldstone mode in the pion-condensed phase, which is valid for momenta $p\ll \mu_I$. It is shown that its damping rate is of order $p^5$ 
in the nonrelativistic limit, which is Beliaev's result for a dilute Bose gas. It is also shown that in the nonrelativistic limit
the energy density can be matched onto the classic result by Lee, Huang and Yang (LHY) for a dilute Bose, with an $s$-wave scattering length that includes radiative corrections.

\end{abstract}

\maketitle

\section{Introduction}

\begin{figure}[htb!]
\centering
\includegraphics[width=8cm]{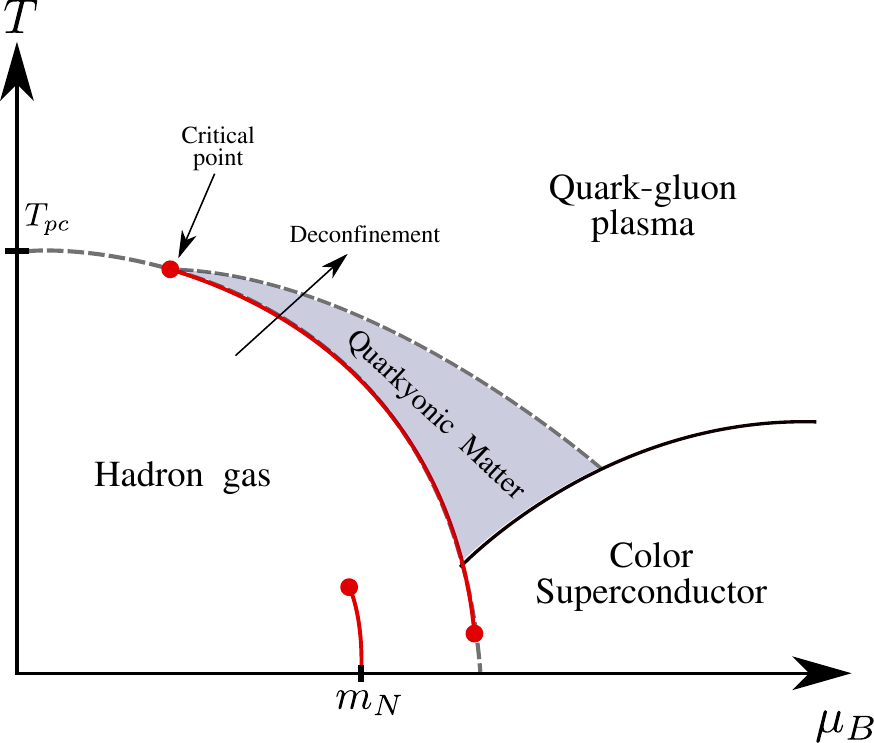}
\caption{Schematic phase diagram of QCD in the $\mu_B$--$T$ plane.
See main text for details.}%
\label{muit}
\end{figure}

The phase diagram of quantum chromodynamics (QCD)
has received a lot of attention in recent years due to its relevance
for the early universe, heavy-ion collisions, and compact stars~\cite{raja,alford,fukurev}. Conventionally, the phase diagram is shown in the plane of temperature $T$ and baryon chemical potential $\mu_B$, see Fig.~\ref{muit}.

At finite $\mu_B$, the sign problem of QCD poses a serious challenge.
The fact that the fermion determinant is complex prohibits the use of importance sampling
techniques in lattice simulations, which precludes the study of a large part of the phase diagram by this method. However, by expanding the partition function in powers of $\mu_B/T$ around zero, one can move away from the temperature axis into the $\mu_B$--$T$ plane, but obviously not too far. The low-$T$, high-$\mu_B$ part of the phase diagram has therefore been mapped out by low-energy models that share some of the properties of QCD, for example the Nambu-Jona-Lasinio (NJL), quark-meson (QM) model, and their Polyakov-loop extended counterparts. This is exactly the part that is relevant for compact stars. Some aspects of the phase diagram are
indicated in Fig.~\ref{muit} in the $\mu_B$--$T$ plane,
for example there may be a quarkyonic phase and a region where QCD is a color superconductor. This part of the phase diagram is 
possibly very rich with a number of phases such as the color-flavor locked (CFL) phase, the two-flavor color superconducting (2SC) phase, and Larkin-Ovchinnikov-Fulde-Ferrel (LOFF) phases. Only the existence of the CFL phase, however, is a rigorous result due to asymptotic freedom. For example the existence of the 2SC phase between the CFL phase and normal quark matter depends heavily on model parameters~\cite{steiner,no2sc,ryster,abuki2}.

Instead of using a common quark chemical potential $\mu={1\over3}\mu_B$ for the different
flavors $f=u,d,s$, one can introduce an independent chemical potential $\mu_f$ for each of them. They are expressed in terms of the baryon, isospin, and
strangeness chemical potentials as $\mu_B={3\over2}(\mu_u+\mu_d)$, $\mu_I=(\mu_u-\mu_d)$,
and $\mu_S={1\over2}(\mu_u+\mu_d-2\mu_s)$.~\footnote{This gives rise to a complicated
three-dimensional phase diagram. In recent years, yet another aspect of the QCD phase diagram has been studied in detail, namely the effects of a strong magnetic background $B$.}
In the special case $\mu_u=-\mu_d\neq0$ and $\mu_s=0$,
only the isospin chemical potential $\mu_I$ is nonzero.
From a theoretical point of view, QCD at zero baryon and strangeness
chemical potentials, but nonzero isospin
chemical potential has the advantage that there is no sign problem: the fermion 
determinant is (manifestly) real and one can use standard importance sampling techniques to perform lattice simulations of the system. This opens up the possibility to compare lattice
with low-energy effective theories such as chiral perturbation theory
and models such as the NJL model and the QM model. Chiral perturbation theory~\cite{wein,gasser1,gasser2}, first considered at finite $\mu_I$ in Refs.~\cite{son,kim,kogut33}, is of particular interest since it gives model-independent
predictions. Within its domain of validity, a comparison with lattice results can also be considered a check of the latter.

\begin{figure}[htb!]
\centering
\includegraphics[width=8cm]{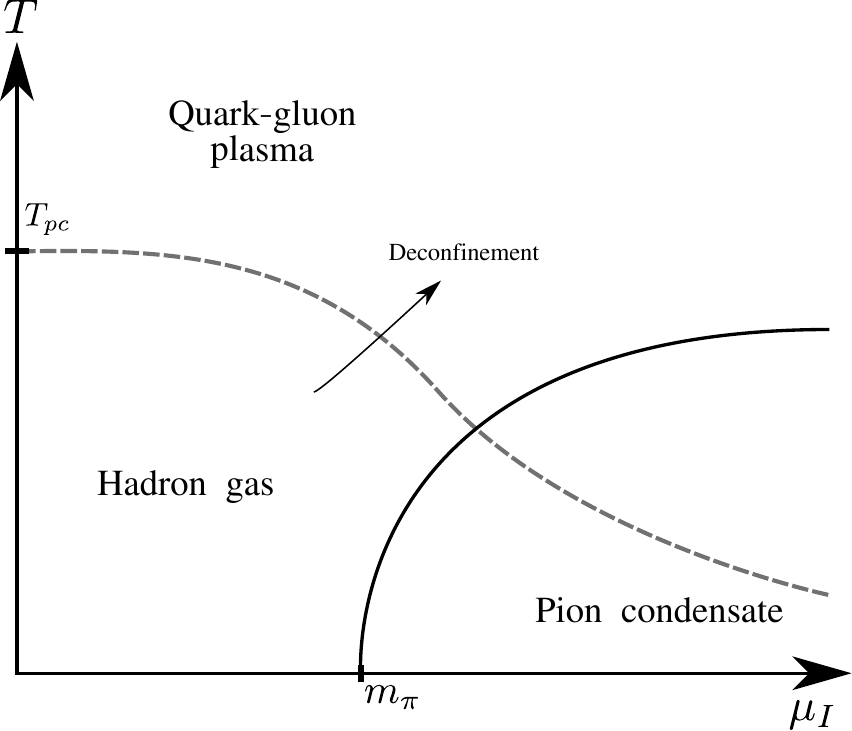}
\caption{Schematic phase diagram of QCD in the $\mu_I$--$T$ plane.
See main text for details.}%
\label{muit2}
\end{figure}

The first simulations of two-flavor QCD at nonzero isospin chemical potential
were performed more than twenty years ago using quenched lattice QCD~\cite{kogut1}, which later was improved by including dynamical fermions~\cite{kogut2,kogut3} on relatively coarse lattices. Later, the simulations were extended to three-flavor QCD in the phase-quenched
approximation~\cite{kogut4,kogut5}. In recent years, high-precision lattice simulations have been
carried out~\cite{gergy1,gergy2,gergy3,gergy4,bcslattice0,bcslattice,newlattice} and the phase diagram in the $\mu_I$--$T$ plane has been mapped out, see  Fig.~\ref{muit2}
(see also Ref.~\cite{verynew} for a determination of the equation of state at finite $\mu_I$). 
The solid black line is the phase boundary between the hadronic phase and the Bose-condensed phase, where the $\mathrm{U}(1)_{I_3}$ symmetry is broken. The grey dashed line is the phase boundary between the confined and the deconfined phases. For large $T$ and small $\mu_I$, this is the usual transition to a quark-gluon plasma. 

It has been shown both on the lattice and in $\chi$PT that the transition at $T=0$
from the vacuum phase to the pion-condensed phase takes place at a critical 
isospin chemical potential equal to the physical charged pion mass, $\mu_I^c=m_{\pi^{\pm}}$
and that the transition is second order. The transition remains second order at finite $T$
along the solid black line. Moreover, the vacuum phase exhibits the so-called 
Silver-Blaze property, meaning that the thermodynamic properties at $T=0$
are independent of the chemical potential $\mu_I$ all the way up to the onset
of Bose-Einstein condensation~\cite{cohen}.
For asymptotically large values of $\mu_I$, quarks rather than pions are the relevant degrees of freedom. Without interactions and {at} low temperature, the system is described in terms of a 
Fermi surface. Turning on interactions, the Fermi surface is rendered unstable due to an attractive interaction channel provided by one-gluon exchange. The system then becomes a color superconductor, which is characterized by a BCS gap~\cite{son}. At lower densities, the interactions are presumably mediated by instantons. The transition from the BEC phase to the BCS phase is a crossover since the quantum numbers of the condensates are the same. This picture is supported by lattice simulations~\cite{bcslattice0,bcslattice} and the crossover is indicated by the dashed line to the right in Fig.~\ref{muit2}. Various aspects of QCD at finite isospin can be found in e.g.
Refs.~\cite{son,kim,kogut33,carig,carig2,usagain,mojahed,condensates,zhou} ($\chi$PT),~\cite{klim1,klim2,harmen,njlisospin1,sidney,tamas} (NJL model),
and~\cite{kovacs,farias2,kojo,newfarias,kojo2} (QM model). Ref.~\cite{gronli} provides an application of  $\chi$PT in an external magnetic field at finite $\mu_I$, and a review on
meson condensation in QCD can be found in Ref.~\cite{mannarellirev}.

Electromagnetic interactions in chiral perturbation theory were first 
included at the leading-order (LO) by Ecker et al.~\cite{ecker} and at the next-to leading-order (NLO) by Urech~\cite{urech1}.
In Ref.~\cite{urech1}, the NLO Lagrangian was derived and the leading corrections 
to the meson masses were computed, i.e. corrections to Dashen's 
theorem~\cite{dashen}. Other early applications of $\chi$PT to e.g. scattering of pions can be found in Refs.~\cite{meissner,urech2,gas2003}. In the context of chiral perturbation theory,  electromagnetic effects have not yet been included at finite chemical potentials.  Since the spontaneously broken $\mathrm{U}(1)_{I_3}$ symmetry group  
in the pion-condensed phase implies the breaking of the local gauge symmetry of electromagnetism, the Goldstone boson disappears from the physical spectrum and the photon becomes massive via the Higgs mechanism. The pion-condensed phase is then a superconductor and QCD is in a Higgs phase.
The same remark applies to the phase of condensed charged kaons, but not to the
phase of condensed neutral kaons. In this case, the phase is a superfluid.

The paper is organized as follows. In 
Sec.~\ref{section: chiral perturbation theory}, we discuss
chiral perturbation theory including chemical potentials as well as 
electromagnetic interactions. In Sec.~\ref{alphaetc}, the relations between
the thermodynamic potential and various thermodynamic quantities are briefly
reviewed. In Sec.~\ref{grfluc}, we discuss the form of the QCD ground state
as a function of the chemical potentials and how to parametrize the fluctuations
around it. The leading-order results for the quasi-particle masses,
thermodynamic quantities, and the phase diagram are presented in Sec.~\ref{lores}.
In Sec.~\ref{nlosec}, we calculate the pressure, isospin density, energy density,
and the speed of sound to the next-to leading order without electromagnetic interactions.
The results are compared with recent lattice simulations. 
We also show how the three-flavor result for the pressure reduces to the
two-flavor result with renormalized couplings.
In Sec.~\ref{chiralimit},
the pressure is calculated to the next-to-next-to leading order (NNLO)
in the chiral limit. The low-energy effective theory for the Goldstone
bosons is discussed in Sec.~\ref{loweft}, where we calculate their 
damping rate in the nonrelativistic limit. In Sec.~\ref{dilutie}, the dilute Bose gas is reviewed. We show
that the nonrelativistic limit of $\chi$PT reproduces the results for 
the dilute Bose gas with a renormalized $s$-wave scattering length.
Finally, we summarize in Sec.~\ref{summary}.

\section{Chiral perturbation theory}
\label{section: chiral perturbation theory}
In this section, we briefly discuss the construction of the low-energy chiral
Lagrangian using the coset construction. This has been done in detail many times in the
literature. Some of the original references are~\cite{leutintro,pichintro}, while
more recent introductions to effective field theory can be found in Refs.~\cite{gripaios,penco,braunerintro}.
The QCD Lagrangian with ${N_f}$ massless quarks has an $\mathrm{SU}(N_{{f}})_L\times \mathrm{SU}(N_{{f}})_R$ global symmetry which in the vacuum is broken down to $\mathrm{SU}(N_{{f}})_V$ by the quark condensate. The vacuum manifold or the coset is then $\mathrm{SU}(N_{{f}})_L\times \mathrm{SU}(N_{{f}})_R/\mathrm{SU}(N_{{f}})_V$. The vacuum manifold of the effective low-energy theory can be parametrized as
\bqa
\label{exponential form}
\Sigma&=&\Sigma_0e^{i\phi_aX_a/f}\;,
\eqa
where $\Sigma_0$ is the vacuum state, $X_a$ are the broken symmetry generators, 
$\phi_a$ are the Goldstone fields, and $f$ is the bare pion decay constant. 
For two flavors, the broken generators are the Pauli matrices $\tau_a$ and in the three-flavor case, the broken generators are the Gell-Mann matrices $\lambda_a$. They are normalized as $\langle\tau_a\tau_b\rangle=\langle\lambda_a\lambda_b\rangle=2\delta_{ab}$,
where $\langle A\rangle$ denotes the trace of the matrix $A$. Under $\mathrm{SU}(N_{{f}})_L\times \mathrm{SU}(N_{{f}})_R$ transformations $L$ and $R$, $\Sigma$ transforms as
\bqa
\Sigma&\rightarrow&\Sigma^{\prime}= 
{L\Sigma R^{\dagger}}\;.
\eqa
The fundamental object or building block in the coset construction
is the so-called Maurer-Cartan form
\bqa
d_{\mu}&=&i\Sigma^{\dagger}\partial_{\mu}\Sigma\;,
\label{maurer}
\eqa
which is an element of the $\mathfrak{su}(N_{{f}})$ algebra. One constructs the invariant terms in the Lagrangian by taking traces of products of Eq.~(\ref{maurer}). Each factor of $d_{\mu}$ counts as one power of momentum via
the derivative of $\Sigma$. The leading term $\langle d_{\mu}\rangle$ clearly vanishes
since the generators of $\mathrm{SU}(N_{{f}})$ are traceless. The next term is the trace of 
a product of two $d_{\mu}$'s with indices contracted so it is Lorentz invariant.
At leading order in the low-energy expansion, the Lagrangian is therefore
\bqa
\label{lolag00}
{\cal L}_2&=&
{1\over4}f^2\langle\partial_{\mu}\Sigma^{\dagger}\partial^{\mu}\Sigma\rangle\;.
\eqa
In real-world QCD, the current quark masses ${m_f}$ are nonzero. This explicit symmetry breaking
in the QCD Lagrangian gives rise to {the symmetry-breaking} terms in the
chiral Lagrangian. Similarly, electroweak interactions also break chiral symmetry.
For example, we cannot rotate a $u$ quark into a $d$ quark since they have different
electric charges. Finally, for each intact global symmetry, we may introduce
an independent chemical potential $\mu_i$ in the QCD Lagrangian. 
However, this can be done simultaneously for $\mu_i$ and $\mu_j$ only if the 
corresponding charges commute.
In QCD, we can introduce an independent chemical potential $\mu_f$ for each quark flavor $f$
{$(f=u,d,s)$}. In order to incorporate these effects in the chiral Lagrangian, it is convenient to couple the QCD Lagrangian to external fields $v_{\mu}$, $a_{\mu}$, $s$, and $p$ as~\cite{gasser1,urech1}
\bqa
\nonumber
{\cal L}_{\rm QCD}&=&
\nonumber
{-{1\over4}G_{\mu\nu}^aG^{\mu\nu}_a}+i\bar{q}\gamma^{\mu}D_{\mu}q+
\bar{q}\gamma^{\mu}\left(v_{\mu}+\gamma^5a_{\mu}\right)q
\\ &&
-\bar{q}(s-i\gamma^5p)q\;.
\eqa
Here, $G_{\mu\nu}^{{a}}$ is the {nonabelian} field-strength tensor 
and $D_\mu=\partial_\mu-i gT_a\mathcal{A}^a_\mu$ is the corresponding covariant derivative where $g$ is the QCD strong coupling constant, $T_a$ are the ${\rm SU(3)_{{c}}}$ generators and 
$\mathcal{A}_\mu^a$ {are} the gluon field{s}. The external fields are then evaluated at {nonvanishing} values corresponding to the symmetry-breaking terms. These are
\bqa
s&=&M={\rm diag}(m_u,m_d,m_s)\;,\\
\label{pseudo}
p&=&-
\lambda_aj\;,\\
l_{\mu}&=&v_{\mu}-a_{\mu}
=\delta_{0\mu}{\rm diag}(\mu_u,\mu_d,\mu_s) + Q_LA_{\mu}\;,\\
r_{\mu}&=&v_{\mu}+a_{\mu}=
\delta_{0\mu}{\rm diag}(\mu_u,\mu_d,\mu_s) + Q_RA_{\mu}
\;,
\eqa
The source $s$ accounts for the quark masses, while $A_{\mu}$ is the
electromagnetic field, and $\mu_f$ accounts for finite density. 
The pseudoscalar source $p$ is needed if we want to calculate a Bose condensate.
We choose $a=2,5$, or $7$ depending on the condensate of interest, $\pi^{\pm}$, $K^{\pm}$
or $K^0/\bar{K}^0$, see Subsec.~\ref{formi}.
Instead of the quark chemical potentials $\mu_f$, we can {introduce}
the baryon, isospin, and strangeness chemical potentials $\mu_B$, $\mu_I$, and $\mu_S$ using
\bqa
\nonumber
{\rm diag}(\mu_u,\mu_d,\mu_s)&=&{1\over3}(\mu_B-\mu_S)\mathbb{1}+{1\over2}\mu_I\lambda_3
+{1\over\sqrt{3}}\mu_S\lambda_8\;,\\ &&
\label{decomp}
\eqa
with 
\bqa
\mu_B&=&{3\over2}(\mu_u+\mu_d)\;,\\
\mu_I&=&\mu_u-\mu_d\;,\\
\mu_S&=&{1\over2}(\mu_u+\mu_d-2\mu_s)\;.
\eqa
The $\mathrm{SU}(3)_L\times \mathrm{SU}(3)_R$ invariance in QCD
can be made local provided the left-handed and right-handed fields $q_L$ and $q_R$, the external fields transform, {and the charge matrices} as
\begin{align}
\label{tr1}
    q_L&\quad \rightarrow\quad Lq_L\;, \\
    q_R&\quad\rightarrow\quad Rq_R\;, \label{tri11}\\
    (s+ip)&\quad\rightarrow\quad R(s+ip)L^{\dagger}\;,\\
    l_{\mu}&\quad \rightarrow\quad Ll_{\mu}L^{\dagger}+iL\partial_{\mu}L^{\dagger}\;, \\
    r_{\mu}&\quad\rightarrow\quad Rr_{\mu}R^{\dagger}+iR\partial_{\mu}R^{\dagger}\;,\\
    Q_L&\quad\rightarrow\quad LQ_LL^{\dagger}\;,\\
    Q_R&\quad\rightarrow\quad RQ_RR^{\dagger}\;.\label{tr2}
\end{align}    
In the effective theory, the local invariance is implemented by replacing 
a partial derivative by a covariant derivative as
\bqa
\label{covar1}
\partial_{\mu}\Sigma&\rightarrow&\nabla_{\mu}\Sigma=\partial_{\mu}\Sigma-ir_{\mu}\Sigma+i\Sigma l_{\mu}\;,\\
\partial_{\mu}\Sigma^{\dagger}&\rightarrow&\nabla_{\mu}\Sigma^{\dagger}=\partial_{\mu}\Sigma^{\dagger}+i\Sigma^{\dagger}r_{\mu}-i l_{\mu}\Sigma^{\dagger}\;.
\label{covar2}
\eqa
One can now use the building blocks $\Sigma$, $\Sigma^{\dagger}$, $\nabla_{\mu}\Sigma$, $\nabla_{\mu}\Sigma^{\dagger}$, and $s \pm ip $ to construct invariant terms. Of course, the external fields do not transform according to Eqs.~(\ref{tr1})--(\ref{tr2}),
once they take the constant values, the symmetries are explicitly broken.
However, by construction, the symmetries in the effective theory are broken in the same way as in QCD. For example, setting 
$Q_R=Q_L= Q=e\,{\rm diag({2\over3},-{1\over3},-{1\over3})}$ 
being the quark charge matrix, ensures that electromagnetic interactions break the flavor symmetry in the correct way.
{As mentioned above, one can introduce as many independent chemical potentials as there are commuting charges~\cite{kapusta,haber}, which in $\chi$PT
is the dimension of the Cartan subalgebra. For $\mathrm{SU}(3)$, the subalgebra consists of $\lambda_3$
and $\lambda_8$ and we can introduce $\mu_I$ and $\mu_S$.}
{The introduction of the chemical potentials via Eq.~(\ref{decomp})
breaks $\mathrm{SU}(3)_V$
down to $\mathrm{U}(1)_{{I_3}}\times \mathrm{U}(1)_{{S}}$. The baryon chemical potential is redundant in $\chi$PT since the mesons have zero baryon charge, $\Sigma$ transforms
trivially under $\mathrm{U}(1)_B$, and the thermodynamic potential is independent of $\mu_B$}.

There are other choices that may be convenient when discussing
kaon condensation in QCD. Eq.~(\ref{decomp}) can be written as
\bqa
\nonumber
{\rm diag}(\mu_u,\mu_d,\mu_s)&=&{1\over3}(\mu_B-\mu_S)\mathbb{1}+{1\over2}\mu_{K^{\pm}}\lambda_Q
+{1\over2}\mu_{K^0}\lambda_K\;,
\\ &&
\eqa
where $\lambda_Q=\lambda_3+{1\over\sqrt{3}}\lambda_8$ and 
$\lambda_{K}=-\lambda_3+{1\over\sqrt{3}}\lambda_8$ and the corresponding chemical potentials
are $\mu_{K^{\pm}}={1\over2}\mu_I+\mu_S$ and 
$\mu_{K^0}=-{1\over2}\mu_I+\mu_S$, respectively.
These are the combinations of $\mu_I$ and $\mu_S$ that correspond to the quark content
of the charged and neutral kaons. When the absolute value of {$\mu_{K^{\pm}}$
or $\mu_{K^0}$} exceeds the corresponding kaon mass, the kaon forms a Bose condensate.

The organization of the effective Lagrangian in a systematic low-energy expansion requires a consistent power-counting scheme. The original scheme of Gasser and Leutwyler~\cite{gasser1,gasser2}, which does not 
{include} electromagnetic {interactions}, is such that a {covariant} derivative counts as ${\cal O}(p)$. The leading derivative term is already given in Eq.~(\ref{lolag00}),
but there is another term {of order $p^2$}, proportional to 
$\langle\Sigma^{\dagger}\chi+\chi^{\dagger}\Sigma\rangle$, where $\chi=2B_0(s + i p)$ and
$B_0$ is a constant related to the quark condensate.
The leading-order Lagrangian is then
\bqa
\label{lolag11}
{\cal L}_2&=&
{1\over4}f^2\langle\nabla_{\mu}\Sigma^{\dagger}\nabla^{\mu}\Sigma\rangle
+{1\over4}f^2\langle\Sigma^{\dagger}\chi+\chi^{\dagger}\Sigma\rangle\;.
\eqa
If we include electromagnetic terms, we need additional counting rules, which were provided
by Urech~\cite{urech1}. Since {the covariant derivative is of order $p$, the term
$QA_{\mu}$ is also of order $p$}.
The charge $e$ and the field $A_{\mu}$ are assigned to be of order
$p$ and order one, respectively.
This rule does not alter the standard chiral counting.
In addition to including $A_{\mu}$ in the covariant derivative above, there is 
a new {invariant} term {of the form}
$\langle Q_L\Sigma Q_R\Sigma^{\dagger}\rangle$~\footnote{This term breaks the flavor symmetries once we set $Q_L=Q_R=Q$, as explained below Eq.~(\ref{covar2}).}
such that the leading-order Lagrangian is
\bqa
\label{electriclolag}
\nonumber
{\cal L}_2&=&-{1\over4}F_{\mu\nu}F^{\mu\nu}
+{1\over4}f^2\langle\nabla_{\mu}\Sigma^{\dagger}\nabla^{\mu}\Sigma\rangle
+{1\over4}f^2\langle\Sigma^{\dagger}\chi+\chi^{\dagger}\Sigma\rangle
\\&&
+C\langle Q\Sigma Q\Sigma^{\dagger}\rangle
+{\cal L}_{\rm gf}+{\cal L}_{\rm ghost}
-eA_{\mu}J^{\mu}_{\rm back}\;,
\eqa
where the first term is the kinetic term for the photons, $C$ is a coupling constant and the gauge-fixing Lagrangian in general $R_{\xi}$ gauge is
\bqa
{\cal L}_{\rm gf}&=&-{1\over2\xi}(\partial_{\mu}A^{\mu}+\xi ef\sin\alpha\phi_1)^2\;,
\label{gflag}
\eqa
where $\xi$ is the gauge parameter. 
Denoting the ghost field by $c$, the ghost Lagrangian is 
\bqa
{\cal L}_{\rm ghost}&=&
\partial_{\mu}\bar{c}\partial^{\mu}c-\xi e^2f^2\sin^2\alpha\bar{c}c+\xi e^2f\sin\alpha\phi_1\bar{c}c\;.
\eqa
Finally, the last term is the coupling of the gauge field to background charges and currents $J^{\mu}_{\rm back}$. This term is necessary to ensure overall electric neutrality~\cite{kapusta}.
We will not need it in the remainder of the paper.

At next-to-leading order, there are many more terms in the chiral Lagrangian. 
For general $\mathrm{SU}(N)$, there are 13 independent terms in addition to contact terms
(contact terms are terms that depend only on the external fields).
For two flavors, there are 10 terms and an additional 14 terms if we include electromagnetism~\cite{gasser1,urech2,meissner}.
For three flavors the number of operators are 12 and 17, respectively~\cite{gasser2,urech1}. 
The reduction in the number of terms for $N_f=2$ and $N_f=3$ is due to the fact that not all
terms in the $\mathrm{SU}(N_f)$ case are linearly independent.
Not all of the terms in the $N_f=2$ and $N_f=3$ case 
are relevant to the present work. Ignoring electromagnetic effects, the operators we need 
in the three-flavor case are
\bqa
{\cal L}_4&=&
 L_1\langle\nabla_{\mu}\Sigma^{\dagger}\nabla^{\mu}\Sigma\rangle^2
\nonumber+L_2\langle\nabla_{\mu}\Sigma^{\dagger}\nabla_{\nu}\Sigma\rangle
    \langle\nabla^{\mu}\Sigma^{\dagger}\nabla^{\nu}\Sigma\rangle
 \nonumber
\\ && \nonumber
    +L_3\langle\nabla_{\mu}\Sigma^{\dagger}\nabla^{\mu}\Sigma
    \nabla_{\nu}\Sigma^{\dagger}\nabla^{\nu}\Sigma\rangle
\\&& \nonumber
+L_4\langle\nabla_{\mu}\Sigma^{\dagger}\nabla^{\mu}\Sigma\rangle
\langle\chi^{\dagger}\Sigma+\chi\Sigma^{\dagger}\rangle
\\ && \nonumber
+L_5\langle\nabla_{\mu}\Sigma^{\dagger}\nabla^{\mu}\Sigma
(\chi^{\dagger}\Sigma+\chi\Sigma^{\dagger})\rangle
\\ &&
\nonumber
+L_6\langle\chi^{\dagger}\Sigma+\chi\Sigma^{\dagger}\rangle^2
+L_7\langle\chi\Sigma-\chi\Sigma^{\dagger}\rangle^2
\\ && 
+L_8\langle\chi^{\dagger}\Sigma \chi^{\dagger}\Sigma
+  \chi\Sigma^{\dagger}\chi\Sigma^{\dagger}\rangle
+H_2\langle\chi\chi^{\dagger}\rangle\;.
\label{l4utan}
\eqa
The parameters $L_1$--$L_{8}$ are low-energy constants, while the parameter $H_2$ is referred to as a high-energy constant. The relations between the bare and renormalized parameters are
\bqa
\label{li}
L_i&=&L_i^r
{-{\Gamma_i(\delta\Lambda)^{-2\epsilon}\over2(4\pi)^2}{1\over\epsilon}}\;,
\\ 
H_i&=&H_i^r
{-{\Delta_i(\delta\Lambda)^{-2\epsilon}\over2(4\pi)^2}{1\over\epsilon}}\;,
\label{hi}
\eqa
where $\Lambda$ is the renormalization scale and $\delta$ is a constant. $\delta=1$ corresponds to minimal subtraction, while $\log \delta=-{1\over2}(\log4\pi-\gamma_E+1)$
corresponds to the modified minimal subtraction scheme~\cite{bijn1}.
The constants 
$\Gamma_i$ and $\Delta_i$ assume the following values~\cite{gasser2}
\begin{align}
    \Gamma_{1}&=\frac{3}{32}\;,&
    \Gamma_{2}&=\frac{3}{16}\;,&\ 
    \Gamma_{3}&=0\;,\\
    \Gamma_{4}&={1\over8}\;,&
    \Gamma_{5}&=\frac{3}{8}\;,&
        \Gamma_{6}&=\frac{11}{144}\;,\\
    \Gamma_{7}&=0\;,&
    \Gamma_{8}&={5\over48}\;,&
    \Delta_{2}&={5\over24}\;.
\end{align}
The renormalized couplings $L_{i}^{r}$ and $H_{i}^{r}$ are scale-dependent and run to ensure the scale independence of observables in $\chi$PT. Since the
bare couplings are independent of $\Lambda$, differentiation of Eqs.~(\ref{li})--(\ref{hi})
with respect to the scale yields the renormalization group equations
\bqa
\Lambda{dL_i^r\over d\Lambda}=
{-{\Gamma_i(\delta\Lambda)^{-2\epsilon}\over(4\pi)^2}}\;,
\hspace{0.2cm}
\Lambda{dH_i^r\over d\Lambda}=
{-{\Delta_i(\delta\Lambda)^{-2\epsilon}\over(4\pi)^2}}\;.
\label{rl}
\eqa
We note that $\Gamma_3=\Gamma_7=0$, which implies that $L_3^r$ and $L_7^r$ do not run, we therefore write $L_3=L_3^r$ and $L_7=L_7^r$. The solutions to the renormalization group equations are
\bqa
L_i^r(\Lambda)&=&
L_i^r(\Lambda_0){+{\Gamma_i\over2(4\pi)^2}{1\over\epsilon}
\left[(\delta\Lambda)^{-2\epsilon}-(\delta\Lambda_0)^{-2\epsilon}\right]}
\;,\\
H_i^r(\Lambda)&=&
H_i^r(\Lambda_0){+{\Delta_i\over2(4\pi)^2}{1\over\epsilon}\left[(\delta\Lambda)^{-2\epsilon}-(\delta\Lambda_0)^{-2\epsilon}\right]}\;,
\eqa
where $\Lambda_0$ is a reference scale. In the limit $\epsilon\rightarrow0$, the solutions reduce to
\bqa
\label{Lrun}
L_i^r(\Lambda)&=&L_i^r(\Lambda_0)-{\Gamma_i\over2(4\pi)^2}\log{\Lambda^{2}\over\Lambda_0^{2}}\;,
\\ 
H_i^r(\Lambda)&=&H_i^r(\Lambda_0)-{\Delta_i\over2(4\pi)^2}\log{\Lambda^{2}\over\Lambda_0^{2}}\;.
\eqa
In the two-flavor case, the relevant terms in NLO Lagrangian are~\footnote{The expression in~\cite{gasser1} is written using $\mathrm{O(4)}$ vector notation, instead of the more common $\mathrm{SU}(2)_L\times\mathrm{SU}(2)_R$ matrix notation employed in this text. 
This leads to some additional numerical factors.
The conversion is explained in Appendix D of~\cite{scherer}.}
\bqa
\nonumber
{\cal L}_4&=&  {1\over4}l_1\langle\nabla_{\mu}\Sigma^{\dagger}\nabla^{\mu}\Sigma\rangle^2
+{1\over4}l_2\langle\nabla_{\mu}\Sigma^{\dagger}\nabla_{\nu}\Sigma\rangle
    \langle \nabla^{\mu}\Sigma^{\dagger}\nabla^{\nu}\Sigma\rangle
\\ && \nonumber
    +{1\over16}(l_3+l_4)\langle\chi^{\dagger}\Sigma+\Sigma^{\dagger}\chi\rangle^2
\\&&
\nonumber
    +{1\over8}{l_4}\langle\nabla_{\mu}\Sigma^{\dagger}\nabla^{\mu}\Sigma\rangle
    \langle\chi^{\dagger}\Sigma+\Sigma^{\dagger}\chi\rangle
    \\ &&
    - {1 \over 16} l_7 \langle\chi^{\dagger}\Sigma-\Sigma^{\dagger}\chi\rangle^2
+{1\over2}h_1\langle\chi^{\dagger}\chi\rangle\;,
\label{l4}
\eqa
where $l_1$--$l_4$, $l_7$, and $h_1$ are bare coupling constants.\footnote{$h_1$ in this text corresponds to  $ h_1 - l_4$ in \cite{gasser1} due to rewriting of terms~\cite{martin}.}
The relations between the bare couplings and their renormalized counterparts are
\bqa
\label{lr1p}
l_i&=&l_i^r
-{\gamma_i(\delta\Lambda)^{-2\epsilon}\over2(4\pi)^2}{1\over\epsilon}\;,\\
h_i&=&h_i^r
-{\delta_i\Lambda^{-2\epsilon}\over2(4\pi)^2}{1\over\epsilon}\;,
\label{lr1}
\eqa
where $\gamma_i$ and $\delta_i$ are pure numbers
\begin{align}
\gamma_1&={1\over3}\;,&
\gamma_2&={2\over3}\;,&
\gamma_3&=-{1\over2}\;,\\ 
\gamma_4&=2\;,&
{\gamma_7}&=0\;,&
\delta_1&=0\;.
\end{align}
The parameters satisfy renormalization group equations similar to Eq.~(\ref{rl}) with solutions that specify their running. For $\epsilon=0$, the solution is
\bqa
\label{lr2}
l_i^r(\Lambda)&=&{\gamma_i\over2(4\pi)^2}\left[\bar{l}_i+\log{m_{\pi,0}^2\over\Lambda^2}\right]\;,
\eqa
where $\bar{l}_i$ are constants. Up to a prefactor these constants equal $l_i^r{(\Lambda)}$
evaluated at the scale of the (bare) pion mass. In the two-flavor case, we will be computing the NNLO pressure in the chiral limit in section~\ref{chiralimit}. In this limit, $m_{\pi,0}\rightarrow0$, so in this case we will use the running coupling $l_i^r(\Lambda)$ instead.

At ${\cal O}(p^6)$, the Lagrangian contains a larger number of terms, 57 for $\mathrm{SU}(2)$ and 94 for $\mathrm{SU}(3)$~\cite{bijn1,bijn2}. Most of them vanish in the chiral limit and for two flavors, the set of operators reduces to 
\bqa
\nonumber
{\cal L}_6&=&
C_{24}\langle(\nabla_{\mu}\Sigma^{\dagger}\nabla^{\mu}\Sigma)^3\rangle
\\ && \nonumber
+C_{25}\langle\nabla_{\rho}\Sigma^{\dagger}\nabla^{\rho}\Sigma\nabla_{\mu}\Sigma^{\dagger}\nabla_{\nu}\Sigma
\nabla^{\mu}\Sigma^{\dagger}\nabla^{\nu}\Sigma
\rangle
\\ &&
+C_{26}\langle\nabla_{\mu}\Sigma^{\dagger}\nabla_{\nu}\Sigma\nabla_{\rho}\Sigma^{\dagger}
\nabla^{\mu}\Sigma\nabla^{\nu}\Sigma^{\dagger}\nabla^{\rho}\Sigma
\rangle\;,
\label{l62}
\eqa
where $C_{24}$--$C_{26}$ are {bare} couplings. The relation between the bare couplings $C_i$ and renormalized couplings $C_i^r$ is 
\bqa
\nonumber
C_{i}&=&
{C_{i}^r(\delta\Lambda)^{-4\epsilon}\over f^2}
-{\gamma_{i}^{(2)}(\delta\Lambda)^{-4\epsilon}\over4(4\pi)^4f^2}{1\over\epsilon^2}
\\ &&
+{(\gamma_{i}^{(1)}+\gamma_{i}^{(L)})(\delta\Lambda)^{-4\epsilon}
\over2(4\pi)^2f^2}{1\over\epsilon}\;,
\label{cjdef}
\eqa
where $\gamma_i^{(1)}$ and $\gamma_i^{(2)}$ are pure numbers. Moreover
\bqa
\gamma_i^{(L)}&=&\sum_j\gamma^{(L)}_{ij}(\delta\Lambda)^{2\epsilon}l_j^r\;,
\eqa
where $\gamma^{(L)}_{ij}$ are also pure numbers.
For $\epsilon=0$, the renormalization group equations for the running couplings
$C_i^r(\Lambda)$ read~\cite{bijn1}
\bqa
\Lambda{dC_i^r\over d\Lambda}&=&
{1\over(4\pi)^2}\left[{2\gamma_i^{(1)}}+{\gamma_i^{(L)}}\right]\;.
\label{rgci}
\eqa
In section~\ref{chiralimit}, we need the combination  $\mathcal{C}=C_{24}+C_{25}+C_{26}$,
cf. Eq.~(\ref{l62}). The relevant coefficients are~\cite{bijn1,bijn2} 
\begin{align}
\gamma_{24}^{(1)}&=-\frac{1}{(4\pi)^{2}}\frac{9}{32}\;,
\gamma_{25}^{(1)}=-\frac{1}{(4\pi)^{2}}\frac{67}{432}\;,\\
\gamma_{26}^{(1)}&=\frac{1}{(4\pi)^{2}}\frac{449}{864}\;,
\gamma_{24}^{(2)}=-{137\over72}\;, \\
\gamma_{25}^{(2)}&={5\over36}\;, 
\gamma_{26}^{(2)}={55\over72}\;, \\
\gamma^{(L)}_{24}&=\left[-2l_1^r-{16\over3}l_2^r-{5\over4}l_6^r\right]
\;,\\
\gamma^{(L)}_{25}&=\left[
2l_1^r-{1\over3}l_2^r +{1\over2}l_6^r\right]\;,\\
\gamma^{(L)}_{26}&= \left[{8\over3}l_2^r +{3\over4}l_6^r
\right]\;.
\end{align}
The renormalized coupling $\mathcal{C}^r=C_{24}^r+C_{25}^r+C_{26}^r$ satisfies the renormalization group
equation
\bqa
\Lambda{d\mathcal{C}^r\over d\Lambda}&=&{1\over6(4\pi)^4}-{3l_2^r\over(4\pi)^2}\;.
\label{dD}
\eqa
The solution to this equation is
\bqa
\nonumber
\mathcal{C}^r(\Lambda)&=&\mathcal{C}^r(\Lambda_0)-{3\over2(4\pi)^2}\log{\Lambda^2\over m_{\pi,0}^2}
\\ && 
\times\left[
l_2^r(\Lambda)
+{1\over6(4\pi)^2}\left(\log{\Lambda^2\over m_{\pi,0}^2}-{1\over3}\right)\right]\;.
\eqa
In the calculation of the thermodynamic quantities in $\chi$PT, we encounter a number of one and two-loop integrals. Some of these integrals are ultraviolet divergent and 
we use dimensional regularization to regulate them.
We introduce the following notation for the integrals in Euclidean space
\bqa
\label{defintegral}
\int_P&=&\int_{-\infty}^{\infty}{dp_0\over2\pi}\int_p\;,
\eqa
where
\bqa
\int_p&=&
\Lambda^{2\epsilon}
\int{d^{d}p\over(2\pi)^{d}}\;,
\label{3ddef}
\eqa
with $P=(p_0,{\bf p})$, $p=|{\bf p}|$ and $d=3-2\epsilon$. The convenience of dimensional regularization is that it automatically sets power divergences
to zero and logarithmic divergences show up as poles in $\epsilon$.
We define the integrals for integers $n\geq0$ as
\bqa
\label{defim2}
I_n(m^2)&=&\int_P{1\over(P^2+m^2)^n}\;,\\
\label{defi0}
I_0^{\prime}(m^2)&=&-\int_P\log\left[P^2+m^2\right]
\;,
\eqa
where the prime denotes differentiation with respect to the index $n$ evaluated at $n=0$.
They satisfy the relations
\bqa
{dI_{0}^{\prime}(m^2)\over dm^2}&=&-I_{1}(m^2)\;,\\
\label{recursion}
{dI_{n}(m^2)\over dm^2}&=&-nI_{n+1}(m^2)\;,
\eqa
which follows directly from the definitions Eqs.~(\ref{defim2})--(\ref{defi0}).
The expression for $I_n(m^2)$ is
\bqa
I_n(m^2)&=&
{m^{4-2n}\over(4\pi)^{{d+1\over2}}}\left({\Lambda\over m}\right)^{2\epsilon}
{\Gamma(n-{d+1\over2})\over\Gamma(n)}\;.
\eqa
The integrals $I_n(m^2)$ are divergent for $n=1$ and $n=2$.
We need the following one-loop integrals expanded to the appropriate order in $\epsilon$
\bqa
\label{i0pp}
I_0^{\prime}(m^2)&=&
{m^4\over2(4\pi)^2}
\left({{\Lambda^{\prime}}\over m}\right)^{2\epsilon}
\left[{1\over\epsilon}+{3\over2}+{\cal O}(\epsilon)\right]\;,\\
I_1(m^2)&=&-{m^2\over(4\pi)^2}\left({{\Lambda^{\prime}}\over m}\right)^{2\epsilon}\bigg[{1\over\epsilon}+1
+{\pi^2+12\over12}\epsilon+  {\cal O}(\epsilon^2)
\bigg]\;,\\
\label{i11}
I_2(m^2)&=&{1\over(4\pi)^2}
\left({{\Lambda^{\prime}}\over m}\right)^{2\epsilon}
\left[{1\over\epsilon}+{\cal O}(\epsilon)\right]\;,
\label{i222}
\eqa
where $\Lambda^{\prime}={\Lambda}(c^2e)^{{1\over2}}$.
The expression for the setting sun diagram is
\bqa
\label{j2def}
J(m^2)&=&
\int_{PQ}{p^2_0\over P^2(Q^2+m^2)[(P+Q)^2+m^2]}\;.
\eqa
Generally, for different nonzero masses, the expression for the setting sun diagram
is complicated. In the present case with
one massless and two equal masses, it simplifies significantly. 
Using Feynman parameters and averaging over angles, it can be written as a product of
two $I_1(m^2)$ as 
\bqa
J(m^2)&=&
{1\over d+1}I_1^2(m^2)\;.
\label{jexp}
\eqa

\section{Thermodynamic potential and thermodynamic quantities}
\label{alphaetc}
The thermodynamic potential $\Omega$ is the object of interest since 
we can calculate quantities such as the pressure, charge densities, and energy density from it. The thermodynamic potential typically depends on particle masses $m_i$,  one or more chemical potentials $\mu_i$
as well as other parameters that we denote by $\alpha_{i}$. 
We have seen that for zero baryon chemical potential, we have two independent chemical potentials e.g. $\mu_I$ and $\mu_S$. However, the phases we discuss in this paper are always described in terms of a single chemical potential, either $\mu_I$ or $\pm{1\over2}\mu_I+\mu_S$. Similarly,
the thermodynamic potential depends on a single additional parameter $\alpha$, which
in QCD can be identified with the rotation angle of the vacuum (see Sec.~\ref{grfluc})
and is nonzero in the Bose-condensed phases. When we discuss the weakly interacting Bose gas, the parameter is denoted by $v$, which is the condensate density.
In the following we therefore write $\Omega(\mu,\alpha)$.

The thermodynamic potential can be written as a low-energy expansion in 
$\chi$PT or as an expansion in powers of
the dimensionless gas parameter $\sqrt{na^3}$
in the case of a dilute Bose gas (See Sec.~\ref{dilute}). Thus we write
\bqa
\Omega(\mu,\alpha)&=&\Omega_0(\mu,\alpha)+\Omega_1(\mu,\alpha)+\cdots+\Omega_m(\mu,\alpha)+\cdots\;,
\eqa
where the subscript $m$ denotes the $m$th-order contribution in the series.
The value of $\alpha$ that extremizes the thermodynamic potential for a given $\mu$ is found by solving the equation
\bqa
\label{dalpha}
{\partial\Omega(\mu,\alpha)\over\partial\alpha}&=&0\;,
\eqa
and is denoted by $\alpha^*$. The pressure ${\cal P}$ is equal to minus the thermodynamic
potential evaluated $\alpha=\alpha^*$, i.e.
\bqa
{\cal P}(\mu)&=&-\Omega(\mu,\alpha^*)\;.
\eqa
The charge density $n$ associated with the chemical potential $\mu$ is
\bqa
n({\mu})&=&-{\partial\Omega(\mu,\alpha)\over\partial\mu}\bigg|_{\alpha=\alpha^*}\;.
\eqa
On the other hand, the pressure ${\cal P}$ is a function of $\mu$ alone,
${\cal P}(\mu)=-\Omega(\mu,\alpha^*(\mu))$. This yields
\bqa
\label{dpdmu}
{d{\cal P}\over d\mu}&=&-\left[{\partial\Omega\over\partial\mu}
+{\partial\Omega\over\partial\alpha^*}{\partial\alpha^*\over\partial\mu}
\right]\;.
\eqa
Since ${\partial\Omega\over\partial\alpha^*}={\partial\Omega\over\partial\alpha}\big|_{\alpha=\alpha^*}$,
the second term in Eq.~(\ref{dpdmu}) vanishes, implying that
\bqa
n(\mu)&=&{d{\cal P}\over d\mu}\;.
\eqa
Finally, the energy density ${\cal E}{(n)}$ is given by a Legendre transform of 
the pressure
\bqa
\label{legendre}
{\cal E}(n)&=&-{\cal P}(\mu)+n(\mu)\mu\;.
\eqa
The solution $\alpha^*$ to Eq.~(\ref{dalpha}) can also written as a series 
\bqa
\alpha^*&=&\alpha_0+\alpha_1+\cdots,
\eqa
where $\alpha_0$ is the LO solution.
Expanding ${\Omega(\mu,\alpha^*)}$ around $\alpha_0$, we obtain
\bqa
\nonumber
{\cal P}(\mu)&=&-\Omega(\mu,\alpha^*)=
-\Omega_0(\mu,\alpha_0)-
{\partial\Omega_0(\mu,\alpha)\over\partial\alpha}\bigg|_{\alpha=\alpha_0}\alpha_1
\\ && \nonumber
-\Omega_1(\mu,\alpha_0)+\cdots\\
&=&-\Omega_0(\mu,\alpha_0 )-\Omega_1(\mu,\alpha_0)+\cdots\;,
\label{ponshell}
\eqa
where we have used that the last term in the first line vanishes, 
cf. Eq.~(\ref{dalpha}).
To NLO in the expansion, the pressure is given by the NLO expression for the thermodynamic potential evaluated at the LO minimum $\alpha_0$.
For completeness, we show how to obtain the first correction $\alpha_1$.
Eq.~(\ref{dalpha}) is expanded as
\begin{align}
\nonumber
0 & =  {\partial\Omega(\mu,\alpha)\over\partial\alpha}\bigg|_{\alpha=\alpha^*}\\
\nonumber
& =
{\partial\Omega_0(\mu,\alpha)\over\partial\alpha}\bigg|_{\alpha=\alpha_0}
+{\partial^2\Omega_0(\mu,\alpha)\over\partial\alpha^2}\bigg|_{\alpha=\alpha_0}\alpha_1
\\ &
+{\partial\Omega_1(\mu,\alpha)\over\partial\alpha}\bigg|_{\alpha=\alpha_0}
+\cdots\;.
\label{expdomega}
\end{align}
Since the first term in the second line of Eq.~(\ref{expdomega}) vanishes, we can easily find the
first correction to $\alpha_0$,
\bqa
\label{alphaein}
\alpha_1&=&
-{\partial\Omega_1(\mu,\alpha)\over\partial\alpha}\bigg|_{\alpha=\alpha_0}\bigg/{\partial^2\Omega_0(\mu,\alpha)\over\partial\alpha^2}\bigg|_{\alpha=\alpha_0}\;.
\eqa

\section{Ground state and fluctuations at finite density}
\label{grfluc}
In this section, we discuss the form of the ground state
and the parametrization of the fluctuations around the ground-state configuration. This discussion has appeared in the literature before, see
for example Ref.~\cite{kim}, but we include it here for completeness.

\subsection{Form of the ground state}
\label{formi}
We first consider two flavors. The QCD vacuum state is $\Sigma_0=\mathbb{1}$. 
In order to determine the ground state at finite isospin, we consider the
most general ${\rm SU}(2)$ matrix with constant fields.
Using Eq.~\ref{exponential form}, we can introduce a real parameter $\alpha$
and constant fields $\hat{\phi}_a$ via $\alpha\hat{\phi}_a=\phi_a/f$, which satisfy $\hat{\phi_a}\hat{\phi}_a=1$. The most general ${\rm SU}(2)$ matrix can then be parametrized as
\bqa
\label{expform}
\Sigma_{\alpha}&=&e^{i\hat{\phi}_a\tau_a\alpha}\;.
\eqa
The subscript $\alpha$ indicates that this parameter characterizes the ground state, which we will show below. Expanding the exponential and using
that the Pauli matrices anticommute, Eq.~(\ref{expform}) can be written as
\bqa
\label{anticom}
\Sigma_{\alpha}&=&\mathbb{1}\cos\alpha+i\hat{\phi}_a\tau_a\sin\alpha\;.
\eqa
The form of the ground state is determined by {minimizing} the classical 
thermodynamic potential $\Omega_0$ as a function of $\hat{\phi}_a$ and $\alpha$, or equivalently by minimizing the static Hamiltonian. The static Hamiltonian is given by $- \mathcal{L}$ evaluated for constant fields. The first term in the Lagrangian Eq.~(\ref{lolag11}) yields the contribution denoted by $\Omega_0^{(1)}({\mu_I,\alpha})$ to the thermodynamic potential~\footnote{In the two-flavor case,
$v_0={1\over3}\mu_B\mathbb{1}+{1\over2}\mu_I\tau_3$, cf. Eq.~(\ref{decomp}).}
\bqa
\nonumber
\Omega_0^{(1)}{(\mu_I,\alpha)}&=&
{1\over4}f^2\langle[v_0,\Sigma^{\dagger}][v_0,\Sigma]\rangle\\
&=&-{1\over2}f^2\mu_I^2(\hat{\phi}_1^2+\hat{\phi}_2^2)\sin^2\alpha\;.
\label{firstterm}
\eqa
We note that this term only depends on the sum of the squares of $\hat{\phi}_1$ and $\hat{\phi}_2$ and is minimized when $\hat{\phi}_1^2+\hat{\phi}_2^2$ is as large as possible,
i.e. when $\hat{\phi}_3$ as small as possible due to the constraint $\hat{\phi}_a\hat{\phi}_a=1$. The second term reads
\bqa
\nonumber
\Omega_0^{(2)}{(\mu_I,\alpha)}&=&
-{1\over4}f^2\langle\chi^{\dagger}\Sigma+\Sigma^{\dagger}\chi\rangle\\
&=&
-f^2B_0(m_u+m_d)\cos\alpha\;.
\label{secondterm}
\eqa
Eq.~(\ref{secondterm}) is independent of $\hat{\phi}_a$. This implies that the thermodynamic potential is minimized for $\hat{\phi}_1^2+\hat{\phi}_2^2=1$.  We now see the competition between the two terms in the Lagrangian:  The first term prefers $\sin\alpha$ as large as possible, and  the second term prefers $\cos\alpha$ as large as possible. Writing $\hat{\phi}_1=\cos\beta$ and 
$\hat{\phi}_2=\sin\beta$, we note that the thermodynamic potential is independent of $\beta$. $\Omega_0(\mu_I,\alpha)$ therefore has a flat direction. We can choose this parameter freely and in the remainder we take $\beta={1\over2}\pi$. The ground state can then be written as
\bqa
\Sigma_{\alpha}&=&
e^{i\tau_2\alpha}=
\mathbb{1}\cos\alpha+i\tau_2\sin\alpha\;.
\label{state}
\eqa
The state Eq.~(\ref{state}) is a rotation of the quark condensate
into a pion condensate by an angle $\alpha$, which characterizes the ground state.

The value of $\alpha$ as a function of $\mu_I$ is now determined by minimizing the leading-order thermodynamic potential
\bqa
\Omega_0(\mu_I,\alpha)&=&
-f^2m_{\pi,0}^2\cos\alpha-{1\over2}f^2\mu_I^2\sin^2\alpha\;,
\label{omegi}
\eqa
where $m^2_{\pi,0}=B_0(m_u+m_d)$ is the physical mass  of the charged pion at tree level (which is degenerate with the neutral pion since we have not included electromagnetic interactions yet, {see section~\ref{lores}}).
The solution $\alpha_0$ as a function of $\mu_I$ satisfies
\bqa
\label{solalpha}
\cos{\alpha_0}=
\left\{ 
\begin{array}{ll}
1\;,  &  \mu_I^2 \leq m_{\pi,0}^2\;, \\
{{m_{\pi,0}^2} \over {\mu_{I }^2 }}\;,
&\mu_I^2 \geq m_{\pi,0}^2\;. \\
\end{array}
\right.
\eqa

We next discuss the parameterization of the ground state in the three-flavor case. The most general ${\rm SU(3)}$ matrix reads 
\bqa
\Sigma_{\alpha}&=&e^{i{\hat{\phi}}_a\lambda_a\alpha}\;.
\eqa
However, since the Gell-Mann matrices do not anticommute, we can not rewrite an ${\rm SU(3)}$ in a convenient form as in the two-flavor case, Eq.~(\ref{anticom}). We can nevertheless use our experience in the ${\rm SU(2)}$ case to write down the form of the ground state.
The hint comes from the three ${\rm SU(2)}$ subgroups of ${\rm SU(3)}$ generated by
$\{\lambda_1,\lambda_2,\lambda_3\}$, 
$\{\lambda_4,\lambda_5,{1\over2}(\lambda_3+\sqrt{3}\lambda_8)\}$,
and $\{\lambda_6,\lambda_7,{1\over2}(\lambda_3-\sqrt{3}\lambda_8)\}$
as well as their associated chemical potentials $\mu_I$, $\mu_{K^{\pm}}$, and $\mu_{K^0}$. Pion condensation involves the first $\rm SU(2)$ subgroup, cf. the Pauli matrices $\tau_a$. Using the same arguments as above, the ground state can be written as
\bqa
\nonumber
\Sigma_{\alpha}^{\pi^{\pm}}&=&
e^{i\lambda_2\alpha}
={1+2\cos\alpha\over3}\mathbb{1}+
{\cos\alpha-1\over\sqrt{3}}\lambda_8
+ i\lambda_2\sin\alpha\;,
\\ &&
\label{pioncond3f}
\eqa
where the superscript indicates the condensing mode. Note that, even though this state is written in terms of $\lambda_8$, which is not a generator of the first ${\rm SU(2)}$ subgroup, this is still an element of that group.
The ground state commutes with $\lambda_3$ only for $\alpha=0$. This means that the $\mathrm{U}(1)_{I_3}$ symmetry is spontaneously broken in the pion-condensed phase, which by Goldstone's theorem leads to a massless excitation. Since the quark charge matrix can be written as ${1\over2}(\lambda_3+{\lambda_8\over\sqrt{3}})$,
the symmetry generated by the electric charge $Q$ is also broken in this phase. In the presence of dynamical photons, this phase is a superconducting Higgs phase with a massive photon. This will discussed in section~\ref{lores}.

Similar arguments apply in the remaining cases of charged and neutral kaon condensation, as the corresponding chemical potential appears together with the generators of the remaining two ${\rm SU(2)}$ subgroups.
The ground states are parametrized as
\bqa
\nonumber
\Sigma_{\alpha}^{K^{\pm}}&=&
e^{i\lambda_5\alpha} 
= {1+2\cos\alpha\over3}\mathbb{1}+
{\cos\alpha-1\over2\sqrt{3}}\left(\sqrt{3}\lambda_3-\lambda_8\right)
\\ &&
+i\lambda_5\sin\alpha
\label{kaon3f}
\;,\\ \nonumber
\Sigma_{\alpha}^{K^0/\bar{K}^0}&=&
e^{i\lambda_7\alpha} 
= {1+2\cos\alpha\over3}\mathbb{1}
+{1-\cos\alpha\over2\sqrt{3}}\left(\sqrt{3}\lambda_3+\lambda_8\right)
\\&&
+i\lambda_7\sin\alpha\;.
\label{kaon3f2}
\eqa
In the case of neutral kaon condensation, the phase is a superfluid, but not a superconductor.
In the next section, we will use the expressions for the ground states to calculate
the pressure and other thermodynamic quantities to leading order in the low-energy
expansion.

The form of our parametrization in the three-flavor case
is based on our experience with the two-flavor case.
For given values of $\mu_I$ and $\mu_S$, we have four candidates for the ground state, namely
the vacuum and the three states in Eqs.~(\ref{pioncond3f})--(\ref{kaon3f2}).
For each of the three states, we minimize the thermodynamic potential. We compare the pressure
of the four candidates and the state with the highest pressure is our ground state.
Strictly speaking, we have not shown that our ground state corresponds to a minimum of the tree-level potential. Expanding in fluctuations about this point, it turns out that the linear terms
vanish (see Subsec.~\ref{fluctu}) and that the masses are real (see Subsec.~\ref{quasi}). Thus
it is locally stable, i.e. at least a metastable point. We will simply assume that it is also the global minimum.

Let us finally comment on the order of the transitions from the vacuum to a Bose-condensed phase. The Bose condensate is an order parameter and in order to calculate it, we need
to couple the system to an external pionic source $j$, cf Eq.~(\ref{pseudo}).
In the case of pion condensation, the external field becomes
\bqa
\chi&=&2B_0M+2iB_0\lambda_2j\;.
\eqa
The leading-order thermodynamic potential then reads
\bqa
\nonumber
\Omega_0(\mu_I,\alpha)&=&-f^2B_0(m_u+m_d){\cos\alpha}-f^2B_0m_s
\\ &&
-2f^2B_0j\sin\alpha-{1\over2}f^2\mu_I^2\sin^2\alpha\;.
\eqa
The quark and pion condensates are then given by the derivatives of $\Omega(\mu_I,\alpha)$ with respect to $m_{u,d}$ and $j$ as 
\bqa
\nonumber
\langle\bar{q}q\rangle&=& 
{{\partial\Omega_0(\mu_I,\alpha)\over\partial m_u}+{\partial\Omega_0(\mu_I,\alpha)\over\partial m_d}}
\\
&=&-2f^2B_0\cos\alpha=\langle\bar{q}q\rangle_{\rm vac}\cos\alpha\;,\\
\nonumber
\langle\bar{q}\gamma^5i\lambda_2q\rangle&=&{\partial\Omega_0(\mu_I,\alpha)\over\partial j}
\\ &=&
-2f^2B_0\sin\alpha=\langle\bar{q}q\rangle_{\rm vac}
\sin\alpha\;,
\eqa
where $\langle\bar{q}q\rangle_{\rm vac}$ is the vacuum value of the quark condensate.
We note that the sum of the square of the condensates is constant and equals the square of the quark condensate in the vacuum. Thus the quark condensate is rotated into a pion condensate.
Using $\cos\alpha=m_{\pi,0}^2/\mu_I^2$, we obtain 
\bqa
{\langle\bar{q}\gamma^5i\lambda_2q\rangle}&=&
\langle\bar{q}q\rangle_{\rm vac}\sqrt{1-{m_{\pi,0}^4\over\mu_I^4}}\;.
\eqa
Close to the phase transition where $\mu_I\approx m_{\pi,0}$ and $\mu_I+m_{\pi,0}\approx 2m_{\pi,0}$, this reduces to
\bqa
\label{close}
{\langle\bar{q}\gamma^5i\lambda_2q\rangle}&=&
2\langle\bar{q}q\rangle_{\rm vac}\sqrt{1-{m_{\pi,0}\over\mu_I}}\;.
\eqa
Thus the order parameter is a continuous function of $\mu_I$ close to the transition, which therefore is of second order. The mean-field critical exponent is ${1\over2}$, which follows directly from Eq.~(\ref{close}) and the system is in the $O(2)$ universality class.

There is another way to see this, based directly on a Ginzburg-Landau analysis of the thermodynamic potential. Expanding the thermodynamic potential Eq.~(\ref{omegi}) in powers of $\alpha$ around $\alpha=0$ to order $\alpha^4$, we obtain
\bqa
\nonumber
\Omega_{{0}}(\mu_I,\alpha)&=&{-f^2m_{\pi,0}^2+}
{1\over2}f^2\left[m_{\pi,0}^2-\mu_I^2\right]\alpha^2\\
\label{rekkje}
&&+{1\over24}f^2\left[4\mu_I^2-m_{\pi,0}^2\right]\alpha^4\;.
\eqa
A critical isospin chemical potential $\mu_I^c$ is defined by the vanishing of the
quadratic term in the equation above. This yields $\mu_I^c=\pm m_{\pi,0}$.
Since the quartic term is positive for $\mu_I=\mu_I^c$, the transition is second order. The value of $\alpha$ that minimizes Eq.~(\ref{rekkje}) is 
\bqa
\alpha_{0}&=&\sqrt{{6(\mu_I^2-m_{\pi,0}^2)\over4\mu_I^2-m_{\pi,0}^2}}\;.
\eqa
Close to the transition, $\alpha=2\sqrt{1-{m_{\pi,0}\over\mu_I}}$.
The onset of BEC when $\mu_I$ is equal to the mass of the charged pion at tree level is a leading-order result, but is expected to hold to all orders 
in the low-energy expansion, where the physical pion mass is calculated in the same approximation as the thermodynamic potential. This was explicitly shown in Ref.~\cite{usagain} to order ${\cal O}(p^4)$, where $m_{\pi,0}$ is replaced by its NLO expression, Eq.~(\ref{nlompi}).

\subsection{Fluctuations}
\label{fluctu}
The ground state $\Sigma_{\alpha}$ minimizes the energy for given values of $\mu_I$ and $\mu_S$. We would like to include quantum corrections and we therefore need to discuss the parametrization of the fluctuations around the ground state. In the vacuum, $\alpha=0$, we have $\Sigma=U\Sigma_0U$, where
\bqa
U&=&
e^{{1\over2}i{\phi_a\lambda_a}/f}\;.
\eqa
A naive way of parametrizing $\Sigma$ for arbitrary $\alpha$ is
\bqa
\label{naive}
\Sigma&=&U\Sigma_{\alpha}U\;.
\eqa
Using this parametrization, one can expand the Lagrangian ${\cal L}_2$ to
second order in the fluctuations. It turns out that the kinetic terms are not
canonically normalized so one needs a field redefinition, which happens to depend 
on the chemical potential. After redefining the field  one can calculate the NLO contribution to the thermodynamic potential arising from the functional determinant in the standard way. 
The counterterms, given by the static part of the NLO Lagrangian ${\cal L}_4$, are
already fixed and do not cancel the divergences
for arbitrary values of $\alpha$, only for the value that corresponds to the
minimum of the tree-level thermodynamic potential, Eq.~(\ref{solalpha}).
In other words, we cannot renormalize the NLO thermodynamic potential away from the
classical minimum and therefore not determine the value of $\alpha$ that 
minimizes it. The problem is that the parametrization Eq.~(\ref{naive})
is not valid for nonzero $\alpha$, which was first pointed out in Ref.~\cite{kim}.
Introducing
\bqa
L_{\alpha}=A_{\alpha}UA_{\alpha}^{\dagger}\;,
\hspace{1cm}
R_{\alpha}=A_{\alpha}^{\dagger}U^{\dagger}A_{\alpha}\;,
\eqa
where $A_{\alpha}=e^{{1\over2}i\lambda_a\alpha}$ (with $a=2,5,7$ depending on the ground state
we consider), the correct parametrization is 
\bqa
\Sigma=L_{\alpha}\Sigma_{\alpha}R^{\dagger}_{\alpha}=
A_{\alpha}(U\Sigma_0U)A_{\alpha}\;.
\label{parameter}
\eqa
Using this parametrization, the kinetic terms are automatically canonically normalized.
Moreover, the ${\cal O}(p^4)$ counterterms cancel the ultraviolet divergences arising
from the functional determinant for all values of $\alpha$, as we shall see in section~\ref{nlosec}.

Using the parametrization Eq.~(\ref{parameter}) in the Lagrangian~(\ref{lolag11})
and expanding in powers of the fields, we obtain
$\mathcal{L}_2= \mathcal{L}_2^{(0)} + \mathcal{L}_2^{(1)}+\cdots$, 
where the superscript indicates the order in the fields.
To be specific, we consider pion condensation.
The zeroth-order term is ${{\cal L}_2^{(0)}=}-\Omega_0{(\mu_I,\alpha)}$, 
from the previous section. The linear term is
\bqa
    \mathcal{L}_2^{(1)} 
&=& f\left[ (\mu_I^2\cos\alpha - m_{\pi,0}^2 ) \phi_2- \mu_I \partial_0 \phi_1\right]\sin\alpha\;.
\eqa
The $\partial_0 \phi_1$ term is a total derivative and can therefore be ignored.
The remaining terms vanish for $\alpha =0 $ or $\cos\alpha = m_{\pi,0}^2 / \mu_I^2$, 
{which exactly is the value of $\alpha$ that minimizes $\Omega_0(\mu_I,\alpha)$.}
In the next section, we calculate the masses showing that the point we expand about is a minimum.
We can draw similar conclusions for the charged and neutral kaon condensates.
For a given pair of $\mu_I$ and $\mu_S$, we calculate the
energy of the ground state of different phases and determine which is the global minimum
and in this manner map out the phase diagram. This is also done in the next section.

We close this section with a few remarks on finite density and symmetry
breaking. The conventional view is that the introduction of a chemical potential $\mu$ breaks Lorentz invariance explicitly. However, in recent years an alternative view has been put forward by Nicolis and 
collaborators~\cite{probing,nicbreak}, namely that Lorentz invariance is broken spontaneously. At finite chemical potential $\mu$, one uses the grand-canonical Hamiltonian defined as
\bqa
{\cal H}^{\prime}&=&{\cal H}-\mu Q\;,
\eqa
where ${\cal H}$ is the original Hamiltonian of the system and $Q$ is the conserved charge that is a consequence of a continuous symmetry via Noether's theorem and 
associated with $\mu$. We are interested in the ground state $|\mu\rangle$
of the modified Hamiltonian ${\cal H}^{\prime}$, satisfying 
\bqa
\label{modi}
{\cal H}^{\prime}|\mu\rangle=0\;.
\eqa
The right-hand side should read $\lambda|\mu\rangle$, but $\lambda$ can always
be set to zero via a redefinition of the cosmological constant~\cite{probing}.
If the charge $Q$ is spontaneously broken, it follows from Eq.~(\ref{modi}) that time translations (which are generated by ${\cal H}$) are also spontaneously broken since the state $|\mu\rangle$ is not an eigenstate of the original Hamiltonian ${\cal H}$.
However, spatial translations are not broken, which singles out the time direction as being special.
The fact that spatial translational symmetries are intact implies that the state $|\mu\rangle$
breaks all Lorentz boosts. On the other hand, a new time translation symmetry is unbroken, namely the translations generated by ${\cal H}^{\prime}={\cal H}-\mu Q$. Summarizing, all the Lorentz boosts and the internal symmetry generated by $Q$ are all spontaneously broken, while translations (space and time), rotations and the internal symmetries not generated by $Q$ remain unbroken. In the case of pion condensation, $Q=Q_{I_3}$ is the third component of the isospin and the internal symmetry that is broken spontaneously by the Bose condensate is 
$\mathrm{U}(1)_{I_3}$.

The idea is that instead of doing perturbation theory around a time-independent ground state, one expands around a time-dependent ground state.
We have denoted the time-independent ground state by $\Sigma_{\alpha}^{\pi^{\pm}}$.
Similarly, we denote the time-dependent ground state by $\Sigma_{\alpha}^{\pi^{\pm}}(t)$, this state can be written as
\bqa
\nonumber
\Sigma_{\alpha}^{\pi^{\pm}}(t)&=&e^{-{1\over2}i\lambda_3\mu_It}\Sigma_{\alpha}^{\pi^{\pm}}
e^{{1\over2}i\lambda_3\mu_It}\\ 
&=&
e^{-{1\over2}i\lambda_3\mu_It}A_{\alpha}\Sigma_{0}A_{\alpha}
e^{{1\over2}i\lambda_3\mu_It}
\;.
\label{lorentz}
\eqa
The ground state Eq.~(\ref{lorentz}) breaks time invariance, hence it is
broken spontaneously.
Perturbation theory is now carried out around this state with the original chiral Lagrangian with $\mu_I=0$, which is Lorentz invariant.
This can be seen as follows. The field, now denoted by $\tilde{\Sigma}$,
is parametrized as
\bqa
\nonumber
\tilde{\Sigma}
&=&
{e^{-{1\over2}i\lambda_3\mu_It}\Sigma e^{{1\over2}i\lambda_3\mu_It}}
\\
&=&
e^{-{1\over2}i\lambda_3\mu_It}A_{\alpha}U\Sigma_{0}UA_{\alpha}e^{{1\over2}i\lambda_3\mu_It}\;.
\eqa
It can then be shown that 
\bqa
\partial_{\mu}\tilde{\Sigma}&=&e^{-{1\over2}i\lambda_3\mu_It}(\nabla_{\mu}\Sigma)
e^{{1\over2}i\lambda_3\mu_It}
\;,\\
\tilde{\Sigma}\chi^{\dagger}&=&e^{-{1\over2}i\lambda_3\mu_It}
\Sigma\chi^{\dagger}e^{{1\over2}i\lambda_3\mu_It}\;,
\eqa
where we have used that $\lambda_3$ and the quark mass matrix commute.
Using $\partial_{\mu}\tilde{\Sigma}$, $\tilde{\Sigma}\chi^{\dagger}$
as well as their Hermitian conjugates as building blocks, together with the cyclicity of the trace, we recover the original Lagrangian.

\section{Leading-order results}
\label{lores}
In this section, we derive some leading-order results, namely the quasiparticle masses in the pion-condensed phase,
pressure, densities (isospin and strangeness), and the phase diagram in the
$\mu_I$--$\mu_S$ plane. However, before these results are presented, we discuss how the parameters in the chiral Lagrangian are related to physical observables. 

\subsection{Parameter fixing}
\label{section: connection to physical observables}
In order to find the tree-level masses, we expand the LO chiral Lagrangian to
second order in the fields,
\bqa
{\cal L}_2^{(2)}
\nonumber
    &=& 
    -{1\over4}F_{\mu\nu}F^{\mu\nu}+
    \frac{1}{2}\partial_\mu \phi_a \partial^\mu \phi_a
    - \frac{1}{2} m_a^2\phi_a^2
    \\ &&
    + \frac{1}{\sqrt{3}} \Delta m^2 \phi_3 \phi_8+\partial_{\mu}\bar{c}\partial^{\mu}c
    -\frac{1}{2\xi}(\partial_\mu A^\mu)^2\;.
    \label{lagfull0}
\eqa
Here the masses are
\bqa
m_1^2&=&B_0(m_u+m_d)+\Delta m^2_{\rm EM}=m_{\pi,0}^2+\Delta m^2_{\rm EM}
\;,\\
m_2^2&=&B_0(m_u+m_d)+\Delta m^2_{\rm EM}=m_{\pi,0}^2+\Delta m^2_{\rm EM}\;,\\
m_3^2&=&B_0(m_u+m_d)=m_{\pi,0}^2\;,\\
m_4^2&=&B_0(m_u+m_s)+\Delta m^2_{\rm EM}=m_{K^{\pm},0}^2+\Delta m^2_{\rm EM}\;,\\
m_5^2&=&B_0(m_u+m_s)+\Delta m^2_{\rm EM}=m_{K^{\pm},0}^2+\Delta m^2_{\rm EM}\;,\\
m_6^2&=&B_0(m_d+m_s)=m_{K^0,0}^2\;,\\
m_7^2&=&B_0(m_d+m_s)=m_{K^0,0}^2\;,\\
m_8^2&=&{1\over3}B_0(m_u+m_d+4m_s)=m_{\eta,0}^2\;,
\eqa
with
\bqa
\Delta m^2&=&B_0(m_d-m_u)\;,\\
\Delta m^2_{\rm EM}&=&{2Ce^2\over f^2}\;.
\eqa
The photon as well as the massless ghost field decouple in the vacuum phase so we do not discuss them any further. Moreover, away from the isospin limit, the off-diagonal terms in Eq.~(\ref{lagfull0}) lead to mixing. The vacuum masses are given by the poles of the propagator
\begingroup
\allowdisplaybreaks
\begin{align}
\nonumber
    m_{\pi^0/\eta^0}^2     & =
 {1\over3}\left(m_{K^\pm, 0}^2 +m_{K^0,0}^2+ m_{\pi,0}^2 
    \right.
\\ & \left.
\label{first}
    \mp\sqrt{\big(m_{K^\pm, 0}^2 +m_{K^0,0}^2- 2m_{\pi, 0}^2 \big)^2 
    + 3(\Delta m^2)^2}\right)    \;, 
    \\
    m_{\pi^\pm}^2 &= m_{\pi, 0}^2 + \Delta m_{\rm EM}^2\;, 
    \\
    m_{K^0}^2 &= m_{K^\pm, 0}^2 + \Delta m^2\;, 
    \\
    m_{K^\pm}^2 &=m_{K^\pm, 0}^2 + \Delta m_{\rm EM}^2\;.
    \label{last}
\end{align}
\endgroup
The mass splittings between neutral and charge mesons have two sources. The first  arises from the electromagnetic interactions, $\Delta m^2_{\rm EM}={2Ce^2\over f^2}$, which
is the same for pions and kaons. This is Dashen's theorem~\cite{dashen}.
The second source of mass splitting is the mass difference between the $u$ and the $d$ quark, encoded in $\Delta m^{2}$. For the pion this takes a more complicated form, due to the mixing of $\pi^0$ and $\eta$. To leading order in $\Delta m^{2}$, Eq.~(\ref{first}) yields
\bqa
    m_{\pi^0}^2 = m_{\pi,0}^2 
    -\frac{1}{4} \frac{(\Delta m^2)^2}{m_{K^\pm,0}^2-m_{\pi,0}^2}\;.
\eqa
Therefore, $m^2_{\pi^\pm} = m^2_{\pi^0} + \Delta m^2_{\rm EM} + \mathcal{O}\left((\Delta m^2)^{2}\right)$, and the mass splitting is dominated by the electromagnetic contribution. 
For the kaon, on the other hand, the contributions are of the same order,
\bqa
m_{K^\pm}^2 = m_{K^0}^2 - \Delta m^2 + \Delta m_{\rm EM}^2\;,
\eqa
Notice that the corrections pull in opposite directions,
decreasing the absolute value of the mass splitting.

The meson masses given above in Eqs.~(\ref{first})--(\ref{last}) are the poles masses at tree level. The measured values of the meson masses are taken from the Particle Data Group~\cite{pdg}, 
\begin{align}
m_{\pi^0}&=134.98\,{\rm MeV}\;, 
&
m_{\pi^{\pm}}&=139.57\,{\rm MeV}\;,
\\
m_{K^{\pm}}&=493.68\,{\rm MeV}\;,
&
m_{K^0}&=497.61\,{\rm MeV}\;.
\end{align}
Solving Eqs.~(\ref{first})--(\ref{last}) 
numerically with the {experimental} values given above, we obtain
\begin{align}
    m_{\pi,0} &= 135.09 \; {\rm MeV}, &
    m_{K^\pm,0} &= 492.43\; {\rm MeV}\;, \\
    \Delta m^2 &= (71.60\; {\rm MeV})^2 \;,&
    \Delta m_{\rm EM}^2 &= (35.09\; {\rm MeV})^2\;.
\end{align}
Using the values for the pion masses above and the decay constant and electromagnetic coupling from~\cite{pdg},
\begin{align}
    f_{\pi}&=92.07{\rm MeV}\;,&
    e&=0.3028,
\end{align}
we find that the coupling $C$ introduced in Ref.~\cite{ecker} is
\bqa
C={f^2\over2e^2}\Delta m^2_{\rm EM}=5.692\times10^{-5}\,({\rm GeV})^4\;,
\eqa
where we haved used that $f=f_{\pi}$ at tree level.
We note in passing that the constant $C$ can be expressed in terms of the mass of the $\rho$ meson, its decay constant $f_{\rho}$, and $f_{\pi}$ as~\cite{frho}
\bqa
C&=&{3m_{\rho}^2f_{\rho}^2\over2(4\pi)^2}\ln\left({f_{\rho}^2}\over f_{\rho}^2-f_{\pi}^2\right)\;.
\eqa
Using the values $f_\pi = 93.3\, \text{MeV}$, $f_\rho = 154\,\text{MeV}$ and $m_\rho = 770\, \text{MeV}$, Urech~\cite{urech1} 
obtains the numerical value $6.11\times 10^{-5} \, (\text{GeV})^4$.

\subsection{Quasiparticle masses}
\label{quasi}
In order to calculate the quasiparticles masses, we expand the LO chiral Lagrangian to the second order in the fields. We do this in the pion-condensed phase, 
similar results can be obtained for the kaon-condensed phases.
For simplicity, we consider $\mu_S=0$.
The quadratic terms are
\bqa
{\cal L}_2^{(2)}
\nonumber
    &=& 
    -{1\over4}F_{\mu\nu}F^{\mu\nu}
    +{1\over2}
    m_A^2\eta_{\mu\nu}A^{\mu}A^{\nu} +
    \frac{1}{2}\partial_\mu \phi_a \partial^\mu \phi_a
    \\ && \nonumber
    + \frac{1}{2} m_{ab} \phi_a\partial_0\phi_b
    -m_{\phi A}^2\phi_2A^0
    - \frac{1}{2} m_a^2\phi_a^2
    \\ && \nonumber
    -ef\sin\alpha \partial_{\mu}A^{\mu}\phi_1
    + \frac{1}{\sqrt{3}} \Delta m^2 \phi_3 \phi_8
   +\partial_{\mu}\bar{c}\partial^{\mu}c
   \\ &&
   -m_c^2\bar{c}c-\frac{1}{2\xi}(\partial_\mu A^\mu)^2\;,
    \label{lagfull}
\eqa
where the diagonal mass terms are
\begingroup
\allowdisplaybreaks
\begin{align}
    \label{m1}
    m_1^2 &= m_{\pi,0}^2\cos\alpha    - \left(\mu_I^2-\Delta m_{\rm EM}^2\right) \cos^2\alpha
    +\xi e^2f^2\sin^2\alpha    \;,    \\
    \label{m2}
    m_2^2 &=  m_{\pi,0}^2\cos\alpha   - \left(\mu_I^2-\Delta m_{\rm EM}^2\right) \cos2\alpha\;,  \\
    \label{m3}
    m_3^2 &=  m_{\pi,0}^2\cos\alpha 
    + \left(\mu_I^2-\Delta m_{\rm EM}^2\right) \sin^2\alpha\;, \\
\nonumber
    m_4^2     &=m_{K^\pm,0}^2 + {1\over2}m_{\pi,0}^2(\cos\alpha-1)- {1\over4}\mu_I^2\cos2\alpha
    \\ &
    \label{m4}    
    + {1\over2}\Delta m_{\rm EM}^2\cos\alpha(\cos\alpha+1)
    \;, \\ \nonumber
    m_6^2 &= 
    m_{K^0,0}^2  + {1\over2}m_{\pi,0}^2(\cos\alpha-1)
    -{1\over4}\mu_I^2\cos2\alpha\\ &
    \label{m6}
    +{1\over2}\Delta m_{\rm EM}^2\cos\alpha(\cos\alpha-1)    \;, \\
    \label{m8}
    m_8^2 &=  m_{\eta, 0}^2
    + \frac{1}{3} m_{\pi,0}^2(\cos\alpha - 1)\;,   \\
        m_A^2&=e^2f^2\sin^2\alpha\;, \\
    m_c^2&=\xi e^2f^2\sin^2\alpha\;,
    \label{mc2}
\end{align}
\endgroup
with $m_6^2 = m_7^2 $ and $m_4^2 = m_5^2$.
The nonvanishing off-diagonal terms are
\begin{align}
    \label{m12}
    m_{12} & = 2 \mu_I\cos\alpha\;,\\
    \label{m45}
    m_{45} & = \mu_I  \cos\alpha\;, \\
    \label{m76}
    m_{67} & =  -  \mu_I  \cos\alpha\;,
\end{align}
with $m_{ab} = -m_{ba}$.
The coupling between $\phi_2$ and $A^0$ is given by 
\bqa
    m_{\phi A}^2 &= ef\mu_I\sin2\alpha\;.
\eqa
The spectra of the mesons are 
\begin{align}
\nonumber
    E_{\pi^0/\eta}^2 &=
    p^2 +
    \frac{1}{2} 
    \left( m_3^2 + m_8^2 \right)
    \\ &
        \label{E1}
    \mp
    \frac{1}{2\sqrt{3}}  \sqrt{3 \left( m_3^2 - m_8^2 \right)^2 + 4(\Delta m^2)^2}
    \;, \\
    E_{\tilde{\pi}^{\pm}}^2 
    &= p^2 +
    \frac{1}{2}
    \left(m_1^2 + m_2^2 + m_{12}^2 \right)
    \label{E2p}
    \nonumber\\   &
    \mp\frac{1}{2}
    \sqrt{4p^2m_{12}^2 +\left(m_1^2 + m_2^2 + m_{12}^2\right)^2- 4 m_1^2 m_2^2}\;, 
    \\
    \label{E3}
    E_{\tilde{K}^{\pm}}^2&= p^2 + m_4^2 + \frac{1}{2} m_{45}^2 \mp
    \frac{1}{2} m_{45} \sqrt{4p^2 + 4 m_4^2 + m_{45}^2}\;,\\
    \label{E4}
    E_{\tilde{K}^0/\tilde{\bar{K}}^0}^2&= p^2 + m_6^2 + \frac{1}{2} m_{67}^2 \pm
    \frac{1}{2} m_{67} \sqrt{4p^2 + 4 m_6^2 + m_{67}^2}\;.
\end{align}
The modes in Eq.~(\ref{E1}) are identified with the neutral pion and the $\eta$. 
The modes in Eq.~(\ref{E2p}) are linear combinations of $\pi^+$ and $\pi^-$, denoted by a $\tilde{\pi}^{\pm}$. At onset of pion condensation, $\tilde{\pi}^+$ coincides with $\pi^+$
and  $\tilde{\pi}^-$ coincides with $\pi^-$. Similar remarks apply to the kaon modes,
Eqs.~(\ref{E3}) and~(\ref{E4}).
The effective masses of the particle are then given by $m = E(p = 0)$.
In the remainder, we choose the Feynman gauge, $\xi=1$.
The ghost and the photon have the same mass and the ghost interacts with the would-be Goldstone,
cf. Eq.~(\ref{gflag}). The cross term in the gauge-fixing Lagrangian cancels the term
$-ef\sin\alpha\partial^{\mu} A_{\mu}\phi_1$ in Eq.~(\ref{lagfull}).
There are still two mixing terms, namely between $\phi_3$ and $\phi_8$, and between
$\phi_2$ and $A^0$, making the inverse propagator matrix rather complicated.
Due to this, the long expressions for the dispersion relations will be not listed, only the
quasiparticle masses will be given. In the symmetric phase, $\alpha=0$, the mixing
between $\phi_2$ and $A^0$ vanishes and the photon decouples. The ghost and the photon
are both massless. The remaining masses in the symmetric phase are
\begingroup
\allowdisplaybreaks
\begin{align}
    m_{\pi^0/\eta}^2    & =    \frac{1}{3} 
    \bigg(m_{K^\pm, 0}^2 + m_{K^0, 0}^2 + m_{\pi,0}^2
    \label{sym mass pi/eta}
    \\ \nonumber &
    \mp \sqrt{  \big(m_{K^\pm, 0}^2 + m_{K^0, 0}^2 - 2m_{\pi, 0}^2 \big)^2     + 3(\Delta m^2)^2}\bigg)\;, 
    \\
m_{\tilde{\pi}^\pm}^2 &= \left(\sqrt{m_{\pi, 0}^2 + \Delta m_{\rm EM}^2} \mp \mu_I\right)^2\;,
    \label{sym mass pi pm}
    \\
    m_{\tilde{K}^\pm}^2 &= \left(\sqrt{m_{K^\pm, 0}^2 + \Delta m_{\rm EM}^2}\mp \frac{1}{2}\mu_I\right)^2\;,
    \label{sym mass kpm}
    \\
    m_{\tilde{K}^0/\tilde{\bar{K}}^0}^2  &= \left(m_{K^0, 0} \mp \frac{1}{2} \mu_I\right)^2\;.
    \label{sym mass k0}
\end{align}
\endgroup
Due to the finite chemical potential $\mu_I$, the charged excitations Eqs.~(\ref{sym mass pi pm})--(\ref{sym mass kpm}) are linear combinations of the corresponding excitations in the vacuum
and they are denoted by $\tilde{\pi}^{+}$ etc. 

\begin{figure}[htb!]
\centering
\includegraphics[width=.9\columnwidth]{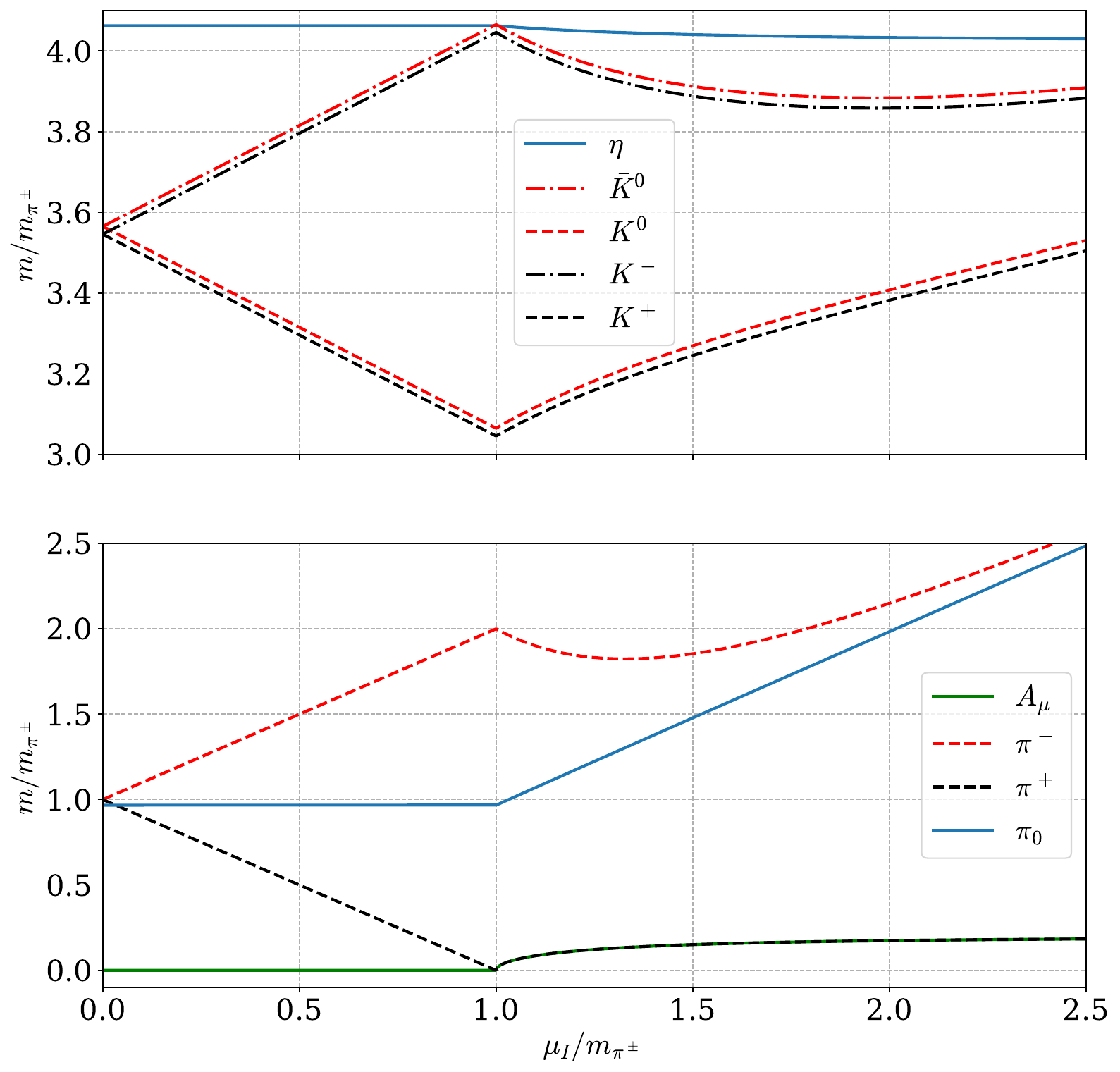}
\caption{Quasi-particle masses at leading order in $\chi$PT as functions of the isospin
chemical potential $\mu_I$ normalized to $m_{\pi^{\pm}}$.
See main text for details.}%
\label{masses higgs}
\end{figure}

In the last section, we found that the transition from the vacuum phase to the symmetry-broken 
phase happens at $\mu_I^2=m_{\pi,0}^2$. When electromagnetic effects are included, the critical
chemical potential is $\mu_{I,\rm eff}^2=\mu_I^2-\Delta m_{\rm EM}^2=m_{\pi,0}^2$ or $\mu_I^2=m_{\pi^{\pm},0}^{2}$ (see details in Subsec.~\ref{pd}).
In the pion-condensed phase, the $\mathrm{U}(1)$ symmetry generated by $\lambda_3$ is broken and 
in the absence of electromagnetic effects, a massless excitation appears in the spectrum.
The generator of electric charge $\lambda_Q = \lambda_3 + \frac{1}{\sqrt 3} \lambda_8$.
Once this symmetry is gauged, it can no longer be broken (Elitzur's theorem), the Goldstone boson disappears from the spectrum, being ``eaten" by the photon via the Higgs mechanism. As a result, the photon becomes massive with three polarization states. At tree level, we have $\cos\alpha_0=m_{\pi,0}^2/\mu_{I,\rm eff}^2={m_{\pi,0}^2/(\mu_{I}^2-\Delta m_{\rm EM}^2)}$,
see Eq.~(\ref{extreme}) below. Defining $\tilde m^2 =\frac{2}{3}(m_{K^\pm,0}^2 + m_{K^0,0}^2- m_{\pi,0}^2)$, the different masses are
\begin{widetext}
\begingroup
\allowdisplaybreaks
\begin{align}
\label{jaja}
    m^2_{\pi^{0}/\eta} &= 
        \frac{1}{2} 
        (\tilde m^2 + \mu_{I,{\rm eff}}^2) + \frac{m_{\pi,0}^4}{6\mu_{I,{\rm eff}}^2} 
        \mp   \frac{1}{6\mu_{I,{\rm eff}}^2}  \sqrt{12 (\Delta m^2)^2 \mu_{I,{\rm eff}}^4 
            + \big[3\mu_{I,{\rm eff}}^2(\tilde m^2 - \mu_{I,{\rm eff}}^2)  + m_{\pi,0}^4\big]^2}\;,
        \\ \nonumber
    m^2_{\tilde{K}^0/\tilde{\bar{K^0}}} &=    m_{K^0,0}^2 + \frac{1}{4}\mu_I^2
         + \frac{m_{\pi,0}^2\mu_I^2(m_{\pi,0}^2 - \mu_{I,{\rm eff}}^2)}{2 \mu_{I,{\rm eff}}^4}  \\
        & \quad
        \mp \frac{m_{\pi, 0}^2|\mu_I|}{2 \mu_{I,{\rm eff}}^4}
        \sqrt{\mu_I^2 m_{\pi,0}^4 +  \mu_{I,{\rm eff}}^2\big[
        \mu_I^2 \big( \mu_{I,{\rm eff}}^2 - 2m_{\pi,0}^2 \big) 
        + 4 m_{K^0,0}^2\mu_{I,{\rm eff}}^2     \big]}    \;,   \\\nonumber
    m^2_{\tilde{K}^\pm} &= m_{K^\pm,0}^2 + \frac{1}{4} \mu_I^2 +
     \frac{m_{\pi, 0}^2}{2\mu_{I,{\rm eff}}^4}\big[
        \mu_I^2 m_{\pi, 0}^2 - \mu_{I,{\rm eff}}^2 \big( \mu_{I,{\rm eff}}^2 - \Delta m_{\rm EM}^2 \big) \big]\mp \frac{m_{\pi, 0}^2|\mu_I| }{2 \mu_{I,{\rm eff}}^4}\\& \quad
    \times\sqrt{ \mu_I^2 m_{\pi, 0}^4 + \mu_{I,{\rm eff}}^2\big[
   {\mu_{I}^2} \big( \mu_{I,{\rm eff}}^2 - 2m_{\pi, 0}^2 \big) 
    + 4 m_{K^\pm,0}^2 \mu_{I,{\rm eff}}^2+ 4\Delta m_{\rm EM}^2m_{\pi, 0}^2\big]}\;,\\ 
m_{\pi^-}^2&= \mu_{I, {\rm eff}}^2\left[1 + 
\frac{ m_{\pi,0}^4 \big( 3\mu_{I, {\rm eff}}^2 + 4 \Delta m_{\rm EM}^2 \big)}{\mu_{I, {\rm eff}}^6}\right]\;,\\
    m_A^2 &=    e^2 f^2 \left[ 1 - \frac{m_{\pi,0}^2}{\mu_{I,{\rm eff}}^2}\right]\;.
\end{align}
\endgroup
\end{widetext}
In Fig.~\ref{masses higgs}, 
the masses are plotted as functions of the isospin chemical potential normalized to the
mass of the charged pion.
We see that the masses are continuous functions of the chemical potential, but they are non-differentiable across the phase transition. Note also that the charged pion vanishes
from the spectrum at the critical chemical potential
(green line in the lower panel). The photon becomes massive, as explained above.

\begin{figure}[htb!]
\centering
\includegraphics[width=.9\columnwidth]{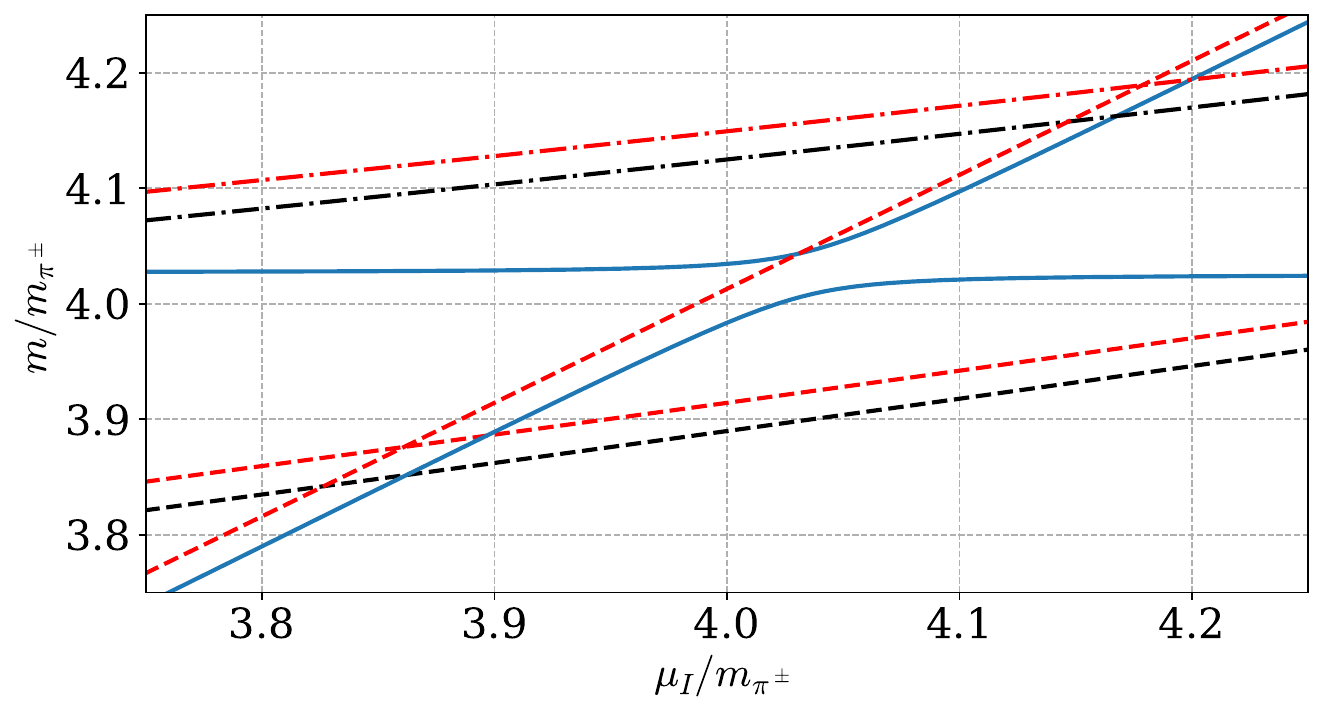}
\caption{The masses of the $\pi^0$ and the $\eta$ approach each other before ``changing roles'', as described in the main text.}%
\label{masses higgs 2}
\end{figure}

Fig.~\ref{masses higgs 2} shows the behavior of the masses for higher values of the isospin chemical potential.
At one point, the mass of the neutral pion and $\eta$ come close to overlapping.
This happens as the second term under the root in Eq.~(\ref{jaja}) vanishes. 
If $\Delta m^2 = 0$, the root vanishes and the absolute value leads to a non-differentiable behavior as the lines intersect.
For $\Delta m^2 \neq 0$, the lines are smooth, and there opens up a gap, as indicated in the figure.
Above this point, the two masses ``change roles'', as the neutral pion mass approaches a constant, while the $\eta$ mass grows linearly.

\subsection{Thermodynamic quantities and phase diagram}
\label{pd}
In this section, we calculate the pressure ${\cal P}$ and
present the phase diagram at $T=0$ in the
$\mu_I$--$\mu_S$ plane for three-flavor $\chi$PT including electromagnetic effects.
The leading-order thermodynamic potential in the pion-condensed phase is
\bqa
\nonumber
\Omega_0(\mu_I, \alpha) &=&
-f^2\left[m_{\pi,0}^2\cos\alpha 
+ B_0 m_s +{1\over3}\Delta m_{\rm EM}^2
\right. \\ &&\left.
+ \frac{1}{2} (\mu_I^2 - \Delta m_{\rm EM}^2)\sin^2\alpha\right]\;,
\eqa
where we have used the form of the ground state $\Sigma_\alpha^{\pi^\pm}$ given in Eq.~(\ref{pioncond3f}). The value $\alpha$ that extremizes $\Omega_0(\mu_I,\alpha)$ satisfies
\bqa
\cos\alpha_0&=&{m_{\pi,0}^2\over\mu_I^2-\Delta m_{\rm EM}^2}={m_{\pi,0}^2\over\mu^2_{I,\rm eff}}\;,
\label{extreme}
\eqa
valid for $\mu_{I,\rm eff}^2=\mu_I^2-\Delta m_{\rm EM}^2\geq m_{\pi,0}^2$. 
The transition therefore takes place at
$\mu_{I,\rm eff}^2=m_{\pi,0}^2$, which is equivalent to $\mu_I^2=m_{\pi^{\pm}}^2 = m_{\pi, 0}^2 + \Delta m_{\rm EM}^2$, i.e. the tree-level mass of the charged pion.
The pressure is expressed as ${\cal P}(\mu_I)=-\Omega_0(\mu_I,\alpha_0)$. Subtracting the pressure in the vacuum phase, we obtain
\bqa
\label{pppion}
{\cal P}={1\over2}f^2\mu_{I,\rm eff}^2\left[1-{m_{\pi,0}^2\over\mu_{I,\rm eff}^2}\right]^2\;.
\eqa
In the pion-condensed phase, the isospin and strangeness densities 
$n_I$ and $n_S$ are obtained from $n = \frac{d \mathcal{P}}{d \mu}$, which gives
\bqa
n_I=f^2\mu_{I}\left[1-{m_{\pi,0}^4\over\mu_{I,\rm eff}^4}\right]\;,
\hspace{0.8cm}
n_S=0\;.
\label{tettleik}
\eqa
We note that the leading-order results for the pion-condensed phase,
Eqs.~(\ref{pppion})--(\ref{tettleik}),
are independent of $m_s$ and are identical to the results in two-flavor $\chi$PT.

The pressure and the densities of the other phases can be calculated in the same way, using the corresponding parametrization of the ground states $\Sigma_\alpha^{K^\pm}$ and $\Sigma_\alpha^{K^0/\bar K^0}$.
In the charged kaon condensed phase, we obtain
\bqa
{\cal P}&=&{1\over2}f^2\mu_{K^{\pm},\rm eff}^2\left[1-{m_{K^{\pm},0}^2\over\mu_{K^{\pm},\rm eff}^2}\right]^2\;,
\\
n_I &=& \frac{1}{2}n_S={1\over2}f^2\mu_{K^\pm}\left[1-{m_{K^{\pm},0}^4\over\mu_{K^{\pm},\rm eff}^4}\right]\;,
\eqa
where  $\mu_{K^{\pm},\rm eff}^2=\mu_{K^{\pm}}^2-\Delta m_{\rm EM}^2$. 
Finally, in the neutral kaon condensed phase, we find
\bqa
{\cal P}&=&{1\over2}f^2\mu_{K^0}^2
\left[1-{m_{K^0,0}^2\over\mu_{K^0}^2} 
\right]^2\;,\\
n_I&=& - \frac{1}{2}n_S=-{1\over2}f^2\mu_{K^0}\left[1-{m_{K^0,0}^4\over\mu_{K^0}^4} \right]\;.
\eqa
In order to find the transition line between two condensed phases, we equate the pressure of them. This gives rise to a line in the phase diagram, which can be solved for
one of the chemical potentials as a function of the other chemical potential and the
vacuum masses of the condensing modes in adjacent phases. 
For example, the line between the charged condensed phases satisfies
\bqa
\nonumber
\mu_{K^{\pm},\rm eff}&=&
\pm{1\over2\mu_{I,\rm eff}}\left(\mu_{I, \rm eff}^2-m_{\pi,0}^2\right.\\ && \left.+
\sqrt{(\mu_{I, \rm eff}^2-m_{\pi,0}^2)^2+4\mu_{I,\rm eff}^2 m_{K^{\pm},0}^2}\right)\;.
\eqa

\begin{figure}[htb!]
\centering
\includegraphics[width=1.\columnwidth]{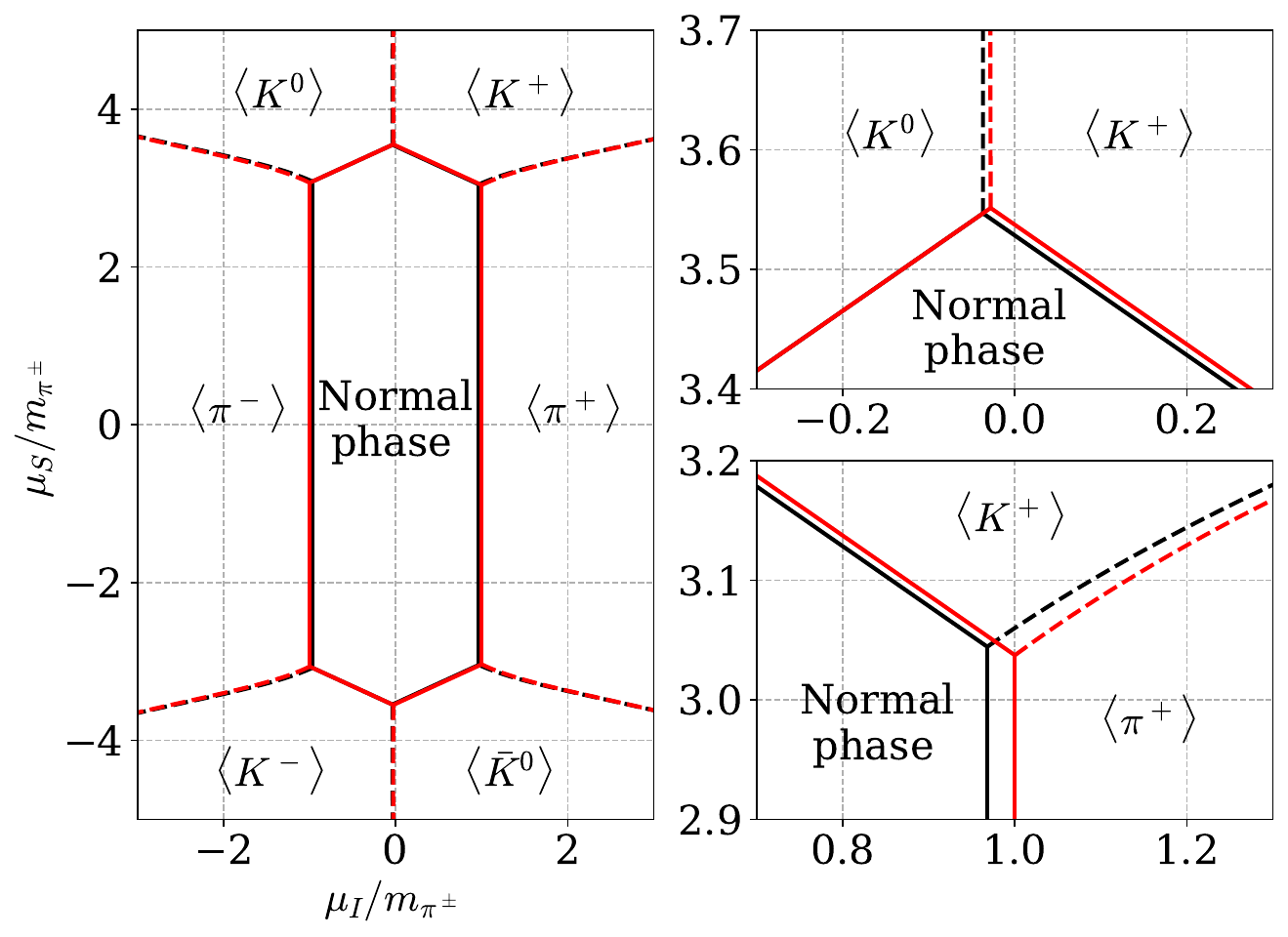}
\caption{Left panel: Tree-level phase diagram in the $\mu_I$--$\mu_S$ plane. Right panel: zoom in on the {triple} points.
See main text for details.}%
\label{phase}
\end{figure}

In Fig.~\ref{phase}, we show the phase diagram in the $\mu_I$--$\mu_S$ plane (left panel) and the details near the triple points
(right panel). The black lines are without electromagnetic effects and in obtaining the red lines they are included. Solid lines separating the phases indicate that the transition is second order, while dotted lines indicate a first-order transition.
In the isospin limit ($m_u=m_d$), this phase diagram was first obtained in Ref.~\cite{kogut33}. 
All the transitions from the vacuum phase are second order, as we have already demonstrated in the transition to the pion-condensed phase.
The transitions are in the $O(2)$ universality class with mean-field exponents. The transitions between the different condensed phases are first order. This can be seen either by a jump in the relevant condensates or by a discontinuity in the relevant rotation angles. 
The points where three phases meet (six points) are the triple points.

In the right panel, we notice the small offset between the black and red lines. Since we normalize the chemical potentials to the mass of the charged pion, the lines indicating the onset of charge meson condensation move upon including electromagnetic effects. The electromagnetic interaction increases the effective mass of the charged particles, and as a result, a higher chemical potential is needed for condensation to occur. For example, the vertical line separating the vacuum and the pion-condensed phase is now at
\bqa
{\mu_I^c\over m_{\pi^0,0}}={m_{\pi^\pm}\over m_{\pi^0,0}}=
\sqrt{1+ \frac{\Delta m_\mathrm{EM}^2}{ m_{\pi^0,0}^2}}\;.
\eqa

Finally, we notice that the normal phase is independent of the chemical potentials, and so the pressure is constant and the densities vanish in the entire region. This is an example
of the so-called Silver Blaze property~\cite{cohen}.
More generally, the properties of a specific Bose-condensed phase depend only on a single chemical potential, namely the relevant one, $\mu_I$, 
$\mu_{K^{\pm}}$, or $\mu_{K^0}$.

\section{Thermodynamics at next-to-leading order}
\label{nlosec}
In this section, we calculate the pressure at next-to-leading order in the low-energy
expansion in the pion-condensed phase. One can obtain similar results for the two kaon-condensed phases. In order to simplify the calculations and to compare our results with lattice simulations, we work in the isospin limit, $m_u=m_d$, and ignore electromagnetic effects.
From the pressure, we calculate other interesting quantities such as the speed of sound, isospin density, and energy density.

\subsection{Thermodynamic potential and pressure}
There are two contributions to the next-to-leading order thermodynamic potential, namely
the static term from ${\cal L}_4$ and the one-loop contribution from ${\cal L}_2$.
The first term reads
\begin{widetext}

\bqa\nonumber
\Omega_1^{(1)}(\mu_I,\alpha)&=&
-(4L_1+4L_2+2L_3)\mu_I^4\sin^4\alpha-8L_4B_0(2m\cos\alpha+m_s)\mu_I^2\sin^2\alpha
-8L_5B_0m\mu_I^2\cos\alpha\sin^2\alpha
\\ &&
-16L_6B_0^2(2m\cos\alpha+m_s)^2-8L_8B_0^2(2m^2\cos2\alpha+m_s^2)-4H_2B_0^2(2m^2+m_s^2)\;,
\label{staticp}
\eqa
where $m = m_u = m_d$. The one-loop contribution from the quasiparticles is given by 
\bqa
\nonumber
\Omega_1^{(2)}(\mu_I,\alpha)&=&
{1\over2}\int_P\log\left[p_0^2+E^{{2}}_{\pi^{0}}(p)\right]
+{{1\over2}}\int_P\log\left[p_0^2+E^{{2}}_{\tilde{\pi}^{\pm}}(p)\right]
+{{1\over2}}\int_P\log\left[p_0^2+E^{{2}}_{\tilde{K}^{\pm}}(p)\right]
\\ &&
+{{1\over2}}\int_P\log\left[p_0^2+E^{{2}}_{\tilde{K}^0/\tilde{\bar{K}}^0}(p)\right]
+{1\over2}\int_P \log\left[p_0^2+E^{{2}}_{\eta}(p)\right]\;,
\label{bare}
\eqa 
where the dispersion relations are given by Eqs.~(\ref{E1})--(\ref{E4}), again with
$\Delta m^2_{\rm EM}=0$,  and we denote the Euclidean four-momentum by $P^2 = p_0^2+ p^2$.
The shorthand notation for the integral is defined in Eqs.~(\ref{defintegral})--(\ref{3ddef}).
The first and the last term in Eq.~(\ref{bare}) are of the form Eq.~(\ref{i0pp}). 
Interestingly, the integrals involving the charged and neutral kaons can also be evaluated directly in dimensional regularization. Combining the two contributions from the charged kaons, we find 
\bqa
\nonumber
\Omega_{1, K^{\pm}}^{(2)}(\mu_I,\alpha)&=&{1\over2}\int_P\log\left[(P^2+m_4^2)(P^2+m_5^2)+p_0^2m_{45}^2\right]\\ 
&=&
{1\over2}\int_P\log\left\{\left[\left(p_0+{im_{45}\over2}\right)^2+p^2+m_4^2+{1\over4}m_{45}^2\right]
\left[\left(p_0-{im_{45}\over2}\right)^2+p^2+m_4^2+{1\over4}m_{45}^2\right]\right\}\;,
\label{ckaons}
\eqa
where we in the second line have used $m_4=m_5$ and factorized.
Shifting the integration variable $p_0$ in the two terms, $p_0\rightarrow p_0\pm {i\over2}m_{45}$,
we obtain
\bqa
\Omega_{1,K^{\pm}}^{(2)}(\mu_I,\alpha)&=&\int_P\log\left[P^2+\tilde{m}_4^2\right]\;,
\label{loopkp}
\eqa
where we have defined $\tilde{m}_4^2=m_4^2+{1\over4}m_{45}^2$. 
The contribution from the neutral kaons is given by the same expression with the 
replacement $m_6^2=m_7^2\rightarrow\tilde{m}_6^2=m_6^2+{1\over4}m_{67}^2$.
Finally, consider the contribution from the charged pions.
We first rewrite their contribution in the same way as for the charged kaons in Eq.~(\ref{ckaons}),
\bqa
\Omega_{1,\pi^{\pm}}^{(2)}(\mu_I,\alpha)&=&{1\over2}\int_P\log\left[(P^2+m_1^2)(P^2+m_2^2)+p_0^2m_{12}^2\right]\;.
\label{cpions}
\eqa
Since $m_1\neq m_2$, we cannot factorize this expression as we did above.
Eq.~(\ref{cpions}) can thus not be evaluated directly in dimensional regularization. 
However, using the techniques from Ref.~\cite{oldnic}, we extract the divergences and express the remainder in terms of a hypergeometric function. Since $m_1=0$ at leading order, the contribution to the pressure is
\bqa\nonumber
\Omega_{1,\pi^{\pm}}^{(2)}(\mu_I,\alpha_{{0}})
&=&{1\over2}\int_P\log\left[P^2(P^2+m_2^2)+p_0^2m_{12}^2\right]
\\
&=&
{1\over2}\int_P\log\left[P^2+m_2^2\right]
-{1\over2}\sum_{n=1}^{\infty}{(-1)^nm_{12}^{2n}\over n}\int_P{p_0^{2n}\over P^{2n}(P^2+m_2^2)^n}\;,
\label{loggi}
\eqa
where we in the second line have expanded the logarithm. Averaging Eq.~(\ref{loggi}) over angles, we find
\bqa
\Omega_{1,\pi^{\pm}}^{(2)}(\mu_I,\alpha_{0})
&=&-{1\over2}I_0^{\prime}(m_2^2)
-{\Gamma(2-\epsilon)\over2\Gamma({1\over2})}\sum_{n=1}^{\infty}{\Gamma(n+{1\over2})\over\Gamma(n+2-\epsilon)}{(-1)^n m_{12}^{2n}\over n}I_n(m_2^2)\;,
\label{divp}
\eqa
where the integrals $I^{\prime}_0(m^2)$ and $I_n(m^2)$ are defined in Eqs.~(\ref{defim2})--(\ref{defi0}). We single out the divergent terms in the series ($n=1,2$) and set $d=3$ in the remainder $(n\geq3)$. Redefining the dummy index $n$ and introducing the Pochhammer symbol $(a)_b=\Gamma[a+b]/\Gamma[a]$, we can write
\bqa
\nonumber
\Omega_{1,\pi^{\pm}}^{(2)}(\mu_I,\alpha_{{0}})
&=&-{1\over2}I_0^{\prime}(m_2^2)+{m^2_{12}\over2(d+1)}I_1(m_2^2)
-{3m^4_{12}\over4(d+1)(d+3)}I_2(m_2^2)
+{5m_{12}^6\over768(4\pi)^2m_2^2}\sum_{n=0}^{\infty}{(1)_n(1)_n({7\over2})_n\over({4})_n({5})_n}{(-{m_{12}^{{2}}\over m_2^{{2}}})^{n}\over n!}\;.
\\ &&
\label{hyper}
\eqa
The expression Eq.~(\ref{hyper}) is identified as the series expansion of a hypergeometric function ${_3F_2}$~\cite{grad}, and we obtain
\bqa
\nonumber
\Omega_{1,\pi^{\pm}}^{(2)}(\mu_I,\alpha_{{0}})&=&
-{1\over2}I_0^{\prime}(m_2^2)+{m^2_{12}\over2(d+1)}I_1(m_2^2)
-{3m^4_{12}\over4(d+1)(d+3)}I_2(m_2^2)
+{5m_{12}^6\over768(4\pi)^2m_2^2}
{_3F_2}
\left[\begin{array}{ccc}
1,&1,&{7\over2}\\
~&4,~5&\\
\end{array}\Bigg|-{m_{12}^2\over m_2^2}\right]\;.
\\ &&
\label{divp222}
\eqa
The hypergeometric function $_3F_2$ has a closed-form expression
\bqa
{_3F_2}\left[\begin{array}{ccc}
1,&1,&{7\over2}\\
~&4,~5&\\
\end{array}\Bigg|z\right]&=&{16\over5}\left[{(3z^2-10z-8)(1-\sqrt{1-z})\over z^4}
+{z^2+4\over z^3}
-3{z^2-4z+8\over z^3}
\log{1+\sqrt{1-z}\over2}\right]\;.
\label{divp2}
\eqa
Adding the contributions, renormalizing the bare couplings according to 
Eqs.~(\ref{li})--(\ref{hi}), and evaluating the expressions for $\cos\alpha_0={m_{\pi,0}^2\over\mu_I^2}$
we obtain the next-to-leading order pressure. In the final result, we subtract the vacuum pressure. This yields
\begingroup
\allowdisplaybreaks
\bqa
\nonumber
{\cal P}_{0+1}&=&
{1\over2}f^2\mu_I^2\left[1-{m_{\pi}^2\over\mu_I^2}\right]
-{1\over2}f^2{m_{\pi,0}^4\over m_{\pi}^2}\left[1-{m_{\pi}^2\over\mu_I^2}\right]
\\ \nonumber&&
+\left[4L_1^r+4L_2^r+2L_3^r+{1\over4(4\pi)^2}\left(
\log{\Lambda^2\over m_2^2}+\log{\Lambda^2\over m_3^2}
+{1\over4}\log{\Lambda^2\over\tilde{m}_4^2}+{9\over8}
\right)\right]\mu_I^4
\\ \nonumber
&&
-\left[32L^r_6+{1\over(4\pi)^2}\left(\log{\Lambda^2\over m^2_{K,0}}+{2\over9}\log{\Lambda^2\over m^2_{\eta,0}}+{11\over18}\right)\right]m^2_{\pi,0}\tilde{m}_{K,0}^2
\\ \nonumber
&&
-\bigg[8L_1^r+8L_2^r+4L_3^r-8L_4^r-4L_5^r+16L^r_6+8L_8^r
\\ \nonumber && 
+{1\over4(4\pi)^2}\left(
3\log{\Lambda^2\over{m}_{\pi,0}^2} +\log{\Lambda^2\over{m_{K,0}^2}}-{1\over2}\log{\Lambda^2\over\tilde{m}_4^2}+{1\over9}\log{\Lambda^2\over m_{\eta,0}^2}+{65\over36}
\right)\bigg]m_{\pi,0}^4
\\ \nonumber&&
+{1\over(4\pi)^2}\left[
\log{{m^2_{K,0}\over\tilde{m}_4^2}}+{4\over9}\log{{m^2_{\eta,0}\over m_8^2}}\right]\tilde{m}_{K,0}^4
\\ \nonumber&&
-\left[8L_4^r-32L_6^r-{1\over2(4\pi)^2}\left(
\log{\Lambda^2\over\tilde{m}_4^2}+{4\over9}\log{\Lambda^2\over m_8^2}+{13\over18}
\right)\right]{m^4_{\pi,0}\tilde{m}_{K,0}^2\over\mu_I^2} \quad 
\\ \nonumber&&
+\bigg[
4L_1^r+4L_2^r+2L_3^r-8L_4^r-4L_5^r+16L_6^r+8L_8^r
\\ && \nonumber
+{1\over144(4\pi)^2}\left(36\log{\Lambda^2\over m_2^2}+9\log{\Lambda^2\over \tilde{m}_4^2}+4\log{\Lambda^2\over m_8^2}-{47\over2}
\right)
\bigg]{m_{\pi,0}^8\over\mu_I^4}
\\ &&  \nonumber
+\left[8L_4^r+
{1\over2(4\pi)^2}\left(
\log{\Lambda^2\over\tilde{m}_4^2}+{1\over2}
\right)
\right]\tilde{m}_{K,0}^2\mu_I^2
\\ &&
-{5m_{\pi,0}^{12}\over12(4\pi)^2(\mu_I^4-m_{\pi,0}^4)\mu_I^4}
{_3F_2}
\left[\begin{array}{ccc}
1,&1,&{7\over2}\\
~&4,~5&\\
\end{array}\Bigg|-{4m_{\pi,0}^4\over \mu_I^4-m_{\pi,0}^4}\right]\;,
\label{nlop}
\eqa
\vspace{4cm}
\end{widetext}
where the masses in Eqs.~(\ref{m2})--(\ref{m4}),~(\ref{m8}), and~(\ref{m45}) are evaluated
at $\cos\alpha_0={m_{\pi,0}^2\over\mu_I^2}$,
\bqa
m_2^2&=&\mu_I^2\left[1-{m_{\pi,0}^4\over\mu_I^4}\right]\;,\\
m_3^2&=&\mu_I^2\;,\\
\tilde{m}_4^2&=&m_4^2+{1\over4}m_{45}^2
= \tilde m_{K,0}^2 
+\frac{1}{4}\mu_I^2\left[1+  \frac{m_{\pi,0}^4}{\mu_I^4}   \right]
\;,\\
\nonumber
 m_8^2  &=&  m_{\eta,0}^2 
- \frac{1}{3}m_{\pi,0}^2 \left[1- \frac{m_{\pi,0}^2}{\mu_I^2}\right]
={1\over3}\left[4\tilde{m}_{K,0}^2+{m_{\pi,0}^4\over\mu_I^2}\right]
\\ &&
\;,
\eqa
\endgroup
with $\tilde{m}_{K,0}^2=B_0m_s$.
In Eq.~(\ref{nlop}), we have subtracted a constant such that the pressure vanishes at $\mu_I=m_{\pi}$. Notice that the explicit $\Lambda$-dependence in Eq.~(\ref{nlop}) cancels against the $\Lambda$-dependence of the renormalized couplings $L_i^r$ as given by Eq.~\ref{Lrun}. This remark about the independence of the renormalization scale applies to all physical quantities.

\subsection{Large $m_s$-mass limit}
The three-flavor pressure is given in Eq.~(\ref{nlop}). In the large-$m_s$ limit,
one expects that the kaons and the eta decouple, and to recover the two-flavor result for the pressure (and all other thermodynamic quantities). The effect of the $s$-quark in this limit is simply to renormalize the couplings, a result one would obtain by integrating it out
at the level of the Lagrangian to obtain a low-energy effective theory for the light mesons. Expanding the three-flavor pressure in inverse powers of $B_0m_s$ and throwing away terms that only depend
on $m_s$, we obtain
\begin{widetext}
\begingroup
\allowdisplaybreaks
\bqa
\nonumber
{\cal P}_{0+1}&=&
{1\over2}\tilde{f}^2\mu_I^2\left[1-{{m}_{\pi}^2\over\mu_I^2}\right]-
{1\over2}\tilde{f}^2{\tilde{m}_{\pi,0}^4\over m_{\pi}^2}\left[1-{{m}_{\pi}^2\over\mu_I^2}\right]
-{m_{\pi,0}^4}\left[2{l}_1^r+2l_2^r
+{l}_3^r+{3\over4(4\pi)^2}
\left(
\log{\Lambda^2\over m_{\pi,0}^2}+{1\over2}
\right)
\right]
\nonumber\\&&
+{m_{\pi,0}^8\over\mu^4_I}\left[{l}_1^r+{l}_2^r+l_3^r
+{1\over4(4\pi)^2}\left(\log{\Lambda^2\mu^2_I\over\mu^4_I-m_{\pi,0}^4}-{5\over6}\right)\right]
+\mu^4_I\left[{l}_1^r+{l}_2^r
+{1\over4(4\pi)^2}\left(\log{\Lambda^{4}\over\mu^4_I-m_{\pi,0}^4}+1\right)\right]
\nonumber\\ &&  
-{5m_{\pi,0}^{12}\over12(4\pi)^2(\mu^4_I-m^4_{\pi,0})\mu^4_I}
{_3F_2}
\left[\begin{array}{ccc}
1,&1,&{7\over2}\\
~&4,~5&
\end{array}\Bigg|-{4m_{\pi,0}^4\over\mu_I^4-m_{\pi,0}^4}\right]\;.
\label{p01final}
\eqa
\endgroup
where we have defined the renormalized parameters
\begingroup
\allowdisplaybreaks
\bqa
\label{tildeb}
\tilde{B}_0m&=&B_0m\left[1-
\left(
16L_4^r-32L_6^r-{2\over9(4\pi)^2}\log{\Lambda^2\over\tilde{m}_{\eta,0}^2}
\right){\tilde{m}_{K,0}^2\over f^2}\right]\;,\\
\label{tildef}
\tilde{f}^2&=&f^2\left[1+\left(
16L_4^r+{1\over(4\pi)^2}\log{\Lambda^2\over\tilde{m}_{K,0}^2}
\right){\tilde{m}_{K,0}^2\over f^2}\right]\;,\\
\label{l1l2def}
l_1^r+l_2^r&=&4L_1^r+4L_2^r+2L_3^r+{1\over16(4\pi)^2}\left[\log{\Lambda^2\over\tilde{m}_{K,0}^2}-1\right]\;,\\
l_3^r&=&-8L_4^r-4L_5^r+16L_6^r+8L_8^r
+{1\over36(4\pi)^2}\left[\log{\Lambda^2\over\tilde{m}_{\eta,0}^2}-1\right]\;,
\label{l3def}
\eqa
\endgroup
with $\tilde{m}_{\pi,0}^2=2\tilde{B}_0m$ and $\tilde{m}_{\eta,0}^2={4\over3}B_0m_s$.
Comparing Eqs.~(\ref{tildeb}) and~(\ref{tildef}) with Eqs.~(\ref{mpi}) and~(\ref{fpi}) below, we see that $\tilde{B_0}m$ and $\tilde{f}$ contain exactly the one-loop corrections
to the pion mass and the pion decay constant from the heavy mesons. Eqs.~(\ref{l1l2def}) 
and~(\ref{l3def}) are in agreement with the result obtained by
Gasser and Leutwyler in Ref.~\cite{gasser2} when comparing two and three-flavor $\chi$PT in the large $m_s$-mass limit. Eq.~\ref{p01final} is in agreement with the recent two-flavor result of Ref.~\cite{zhou} after the identification of $\tilde{f}$ and $\tilde{B}_0m$
as parameters including renormalization effects from $s$-quark loops.

\end{widetext}

\subsection{Numerical results}
\label{numres}
We have expressed our result Eq.~(\ref{nlop}) for the pressure
in terms of the bare masses $m_{\pi,0}$ and $m_{K,0}$, the bare decay constant $f$, the renormalized couplings $L_i^r(\Lambda)$, and the isospin chemical potential $\mu_I$.
In order to evaluate numerically thermodynamic quantities such as the pressure and the energy
density consistently, we need to relate the physical masses (pole masses) to the bare ones. At leading order, this was straightforward as shown in section \ref{section: connection to physical observables}. At next-to-leading order, this requires that we determine these relations also at next-to-leading order. We therefore need the pole masses calculated to one-loop
order. Similarly, the relation between the measured pion decay constant  $f_{\pi}$ and its bare counterpart $f$ receives radiative corrections. The relations are~\cite{gasser2}
\begin{widetext}
\begingroup
\allowdisplaybreaks
  \bqa
  m_{\pi}^2&=&
  m_{\pi,0}^2\left[1
    -\left(8{L}_4^r+8{L}_5^r-16{L}_6^r-16{L}_8^r
      +{1\over2(4\pi)^2}\log{\Lambda^2\over m_{\pi,0}^2}\right)
    {m_{\pi,0}^2\over f^2}
-({L}_4^r-2{L}_6^r){16m_{K,0}^2\over f^2}
  +{m_{\eta,0}^2\over6(4\pi)^2f^2}\log{\Lambda^2\over m_{\eta,0}^2}
  \right]\;,
  \label{mpi}
  \\
m_{K}^2&=&m_{K,0}^2\left[1
  -\left({L}_4^r-2{L}_6^r\right){8m_{\pi,0}^2\over f^2}
    -(2L_4^r+{L}_5^r-4L_6^r-2{L}_8^r){8m_{K,0}^2\over f^2}
-{m_{\eta,0}^2\over3(4\pi)^2f^2}
  \log{\Lambda^2\over m_{\eta,0}^2}\right]\;,
\label{mk}
\\
f_{\pi}^2
&=&f^2\left[1  +\left(8{L}_4^r+8L_5^r+{2\over(4\pi)^2}\log{\Lambda^2\over m_{\pi,0}^2}
\right){m_{\pi,0}^2\over f^2}
+\left(16L_4^r+{1\over(4\pi)^2}\log{\Lambda^2\over m_{K,0}^2}
  \right){m_{K,0}^2\over f^2}\right]\;.
\label{fpi}
\eqa
\endgroup
\end{widetext}
Since we are working in the isospin limit
and with $e = 0$ limit, we only need two physical masses, in contrast to the four used in \ref{section: connection to physical observables}.
In addition, we need the physical value of $f_\pi$ and
the experimental values for the renormalized couplings at a certain scale. 
For three flavors, the convention is that the running couplings are measured at the scale
$\Lambda=m_\rho=770$ MeV.
The couplings needed are listed in 
Table~\ref{table: coupling constants}, taken from 
Ref.~\cite{bijnensMesonicLowEnergyConstants2014}.

\begin{table}
    \centering
    \def\arraystretch{1.2}
    \begin{tabular}{c c c}
        \hline \hline
        Constant & Value [$\times 10^{-3}$] & Source \\
        \hline
        $L_1^r(\Lambda)$ & $\phantom{-}1.0 \pm 0.1 $ & \cite{bijnensMesonicLowEnergyConstants2014} \\
        $L_2^r(\Lambda)$ & $\phantom{-}1.6 \pm 0.2 $ & \cite{bijnensMesonicLowEnergyConstants2014} \\
        $L_3^r(\Lambda)$ & $-3.8 \pm 0.3 $ & \cite{bijnensMesonicLowEnergyConstants2014} \\
        $L_4^r(\Lambda)$ & $\phantom{-}0.0 \pm 0.3 $ & \cite{bijnensMesonicLowEnergyConstants2014} \\
        $L_5^r(\Lambda)$ & $\phantom{-}1.2 \pm 0.1 $ & \cite{bijnensMesonicLowEnergyConstants2014} \\
        $L_6^r(\Lambda)$ & $\phantom{-}0.0 \pm 0.4 $ & \cite{bijnensMesonicLowEnergyConstants2014} \\
        $L_8^r(\Lambda)$ & $\phantom{-}0.5 \pm 0.2 $ & \cite{bijnensMesonicLowEnergyConstants2014} \\
        \hline
    \end{tabular}
\caption{The renormalized coupling constants $L_i^r(\Lambda)$  
of the next-to-leading order
Lagrangian of three-flavor chiral perturbation theory, measured at the scale of the
$\rho$ meson, $\Lambda=m_{\rho}$.}
    \label{table: coupling constants}
\end{table}
In Table~\ref{table: nlo values}, we show the LO and NLO values for the bare parameters.
At LO the bare parameters are equal to the experimental values as explained.
At NLO, they are obtained by solving Eqs.~(\ref{mpi})--(\ref{fpi}) numerically 
using the physical values for $m_{\pi}$, $m_{K}$, $f_{\pi}$, and $L_i^r$ as input.
Since we will be comparing our results with recent lattice simulations~\cite{newlattice},
we use their values for the masses and the pion decay constant and not the values tabulated
by the Particle Data Group~\cite{pdg}.
The values are $m_{\pi}=135.0$ MeV and $m_K=495.0$ MeV. The simulations are carried out with
two different lattices, $24^3\times32$ and $32^3\times48$ and with $f_{\pi}={130\over\sqrt{2}}$
and $f_{\pi}={136\over\sqrt{2}}$, respectively. We choose $f_{\pi}={133\over\sqrt{2}}$
as reasonable value.
At NLO, the pion-decay constant and the mass of kaon receive significant radiative corrections, while the pion mass is hardly affected.
\begin{table}[htb!]
    \centering
    \begin{tabular}{c c c c}
        \hline \hline
        Bare parameter &  LO [MeV] & NLO [MeV] & 1 - LO/NLO \\
        \hline
        $m_{\pi,0}$ & 135.0 &  135.5 & 0.004  \\
        $m_{K,0}$ & 495.0 & 529.4& 0.0649 \\
        $f$ & 133/$\sqrt{2}\approx 94.0$ & 80.8& -0.164\\
        \hline
    \end{tabular}
        \caption{Leading order and next-to-leading order values for the bare masses and decay constant. The values are for $\Delta m^2 = 0$ and $\Delta m_\mathrm{EM}^2 = 0$.
        The physical values, equal to the LO values, are those used in lattice 
        simulations~\cite{newlattice}.}
    \label{table: nlo values}
\end{table}

\begin{widetext}

\begin{figure}[htb]
\includegraphics[width=18cm]{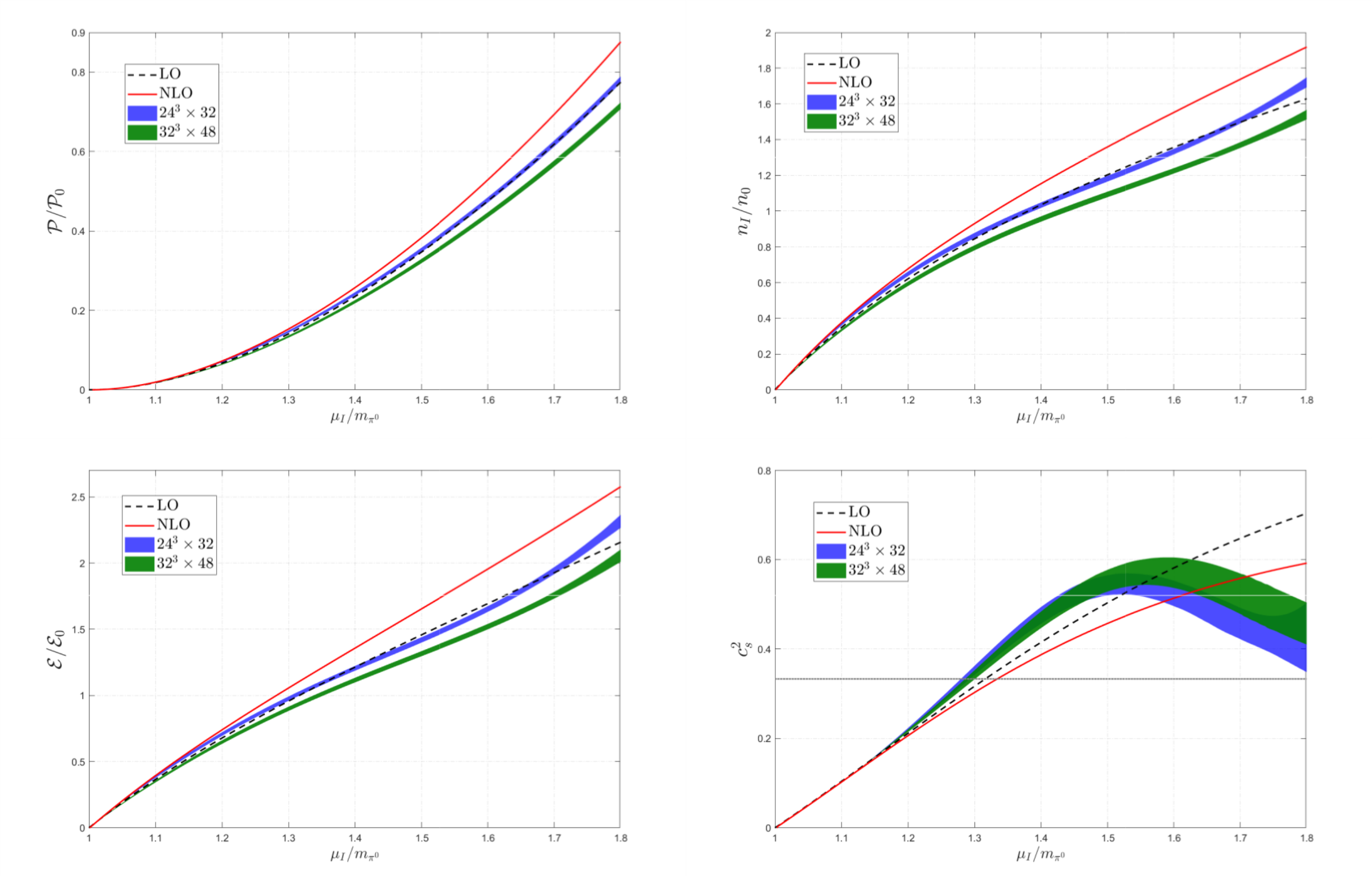}
\caption{LO and NLO results for the normalized pressure, isospin density,
energy density, and speed of sound squared, all functions of
the normalized isospin chemical potential $\mu_I/m_{\pi}$, where $m_{\pi}$ is the physical pion mass at the same order as the approximation.
See main text for details.}
\label{nlores}
\end{figure}

\end{widetext}

In the upper left panel of Fig.~\ref{nlores}, we show the LO (black dashed line)
and NLO (red solid line) results for the pressure ${\cal P}$ normalized to 
${\cal P}_0=f_{\pi}^2m_{\pi}^2$ as a function of $\mu_I/m_{\pi}$.
In the remaining panels, we show the LO and NLO results for the normalized isospin density (upper right panel), normalized energy density (lower left panel), and speed of sound squared $c_s^2$ (lower right panel) as functions of $\mu_I/m_{\pi}$,
with the same color coding. The energy density is normalized to 
${\cal E}_0=f_{\pi}^2m_{\pi}^2$ and the isospin density to $n_0=f_{\pi}^2m_{\pi}$.
The horizontal grey line in the lower right panel is the speed of sound in the conformal
limit, $c_s={1\over\sqrt{3}}$.
Note that since we have normalized the isospin chemical potential
to the physical pion mass determined at the same order as the approximation for the quantity in question, the transition takes place exactly at  $\mu_I/m_{\pi}=1$. The correction going from LO to NLO is increasing as we increase $\mu_I$ and is rather modest for values up to ${\mu_{I}\approx1.3 m_{\pi}}$. For the sake of numerical evaluation, all NLO contributions in {Eq}.~(\ref{nlop}) have been evaluated at the physical masses. That is, from line two and below, we make the substitution $m_{\pi, 0}\rightarrow m_\pi$, and so on. This is consistent to the order in $\chi$PT we are working.

We also compare our results with two sets of lattice data {from Ref.~\cite{newlattice}}. 
The simulations are done on a $24^3\times32$ (blue) and {a}
$32^3\times48$ lattice (green), respectively, where the bands indicate the errors.
We remark in passing that in  Ref.~\cite{newlattice},
the pressure and energy density are normalized to $m_{\pi}^4$ instead of 
$f_{\pi}^2m_{\pi}^2$ and the plots therefore look somewhat different.
The predictions from $\chi$PT are in good agreement for values of $\mu_I$ up to approximately 1.3$m_{\pi}$, somewhat dependent on the quantity in question. Generally, in the region where the NLO results start to deviate from the LO results, the latter is in better agreement with Monte Carlo simulations. This is in contrast to the {predictions} for the quark and pion condensates~\cite{condensates}, here  the NLO {results are} in significantly better agreement with lattice results.

The plot of the speed of sound squared is perhaps particularly interesting. The prediction from $\chi$PT is in good agreement with lattice {for $\mu_I$} up to values of perhaps 1.3{$m_{\pi}$}, whereafter {$\chi$PT} fails badly: The speed of sound increases, as we increase the chemical potential, while the lattice results show a peak around {$\mu_I=1.55m_{\pi}$}, after which it decreases. $\chi$PT is described by an EoS that in the ultrarelativistic limit is $\epsilon=p$
and therefore the speed of sound approaches the speed of light as $\mu_I\rightarrow\infty$.
In contrast to this, in lattice QCD, the relevant degrees at high isospin density are fermions.
The EoS of an ideal Fermi gas is $\epsilon=3p$ and the speed of sound is $c_s={1\over\sqrt{3}}$, which is the conformal limit. Due to asymptotic freedom, it is excepted that QCD approaches this limit as the density increases and the strong interaction gets weaker. Perturbative QCD can be applied for very large values
of $\mu_I$. In a recent paper~\cite{kenji}, the authors use the 
Cornwall-Jackiw-Tomboulis (CJT) formalism~\cite{cjt} to study the
behavior of $c_s$ for large $\mu_I$. The speed of sound does approach 
the conformal limit, but the way depends on whether one includes the BCS gap or not: Including the gap $c_s$ approaches ${1\over\sqrt{3}}$ from above, otherwise $c_s$ approaches the conformal limit from below.

\section{Chiral limit of two-flavor $\chi$PT}
\label{chiralimit}
In two-flavor $\chi$PT at finite $\mu_I$, there are three mass scales, namely the pion mass,
the pion decay constant, and the isospin chemical potential. In the chiral limit, we have only
two mass scales $f$ and $\mu_I$, and a single dimensionless ratio, namely $\mu_I/f$.
All thermodynamic quantities can therefore be expanded in this ratio, and higher-order calculations
are in fact tractable. The main reason is that the propagator is diagonal and has a simple form
due to the fact that $\alpha={1\over2}\pi$. It then follows that the LO pressure is
${\cal P}_0={1\over2}f^2\mu_I^2$ and the pressure can be written as a series
\bqa
{\cal P}_{0+1+2+...}&=&{1\over2}f^2\mu_I^2+a_1\mu_I^4+a_2{\mu_I^6\over f^2}+...\;,
\eqa
where $a_i$ ($i=1,2...$) are coefficients. The pressure through ${\cal O}(p^4)$ in the chiral limit can be found by sending the bare pion mass $m_{\pi,0}$ to zero in Eq.~(\ref{p01final})
and ignoring the loop correction from the $s$-quark to $f$ in Eq.~(\ref{tildef}) Noting that the contribution from the hypergeometric function vanishes in this limit, we obtain the coefficient $a_1$
\bqa
a_1&=&{l}_1^r(\Lambda)+{l}_2^r(\Lambda)
+{1\over2(4\pi)^2}
\left[\log{\Lambda^{2}\over\mu_I^2}+{1\over2}\right]\;.
\label{a1}
\eqa
The scale dependence of the running couplings $l_1^r(\Lambda)$ and $l_2^r(\Lambda)$ in 
Eq.~(\ref{a1}) cancels against the explicit scale dependence of $\log{\Lambda^2\over\mu_I^2}$
in such a way that $a_1$ is independent of $\Lambda$. This remark also applies to $a_2, a_3..$,
and ensures the scale independence of the pressure order by order in the low-energy expansion.

The cubic and quartic interactions from the LO Lagrangian are
\begin{widetext}
\bqa
{\cal L}_2^{\rm (3)}
&=&
{\mu_I\over f}\partial_0\phi_1\left[\phi_2^2+\phi_3^2
\right]\;,\\
{\cal L}_2^{\rm (4)}&=&
{1\over6f^2}\left[\phi_a\phi_b(\partial_{\mu}\phi_a)(\partial^{\mu}\phi_{b})
-\phi_a\phi_a(\partial_{\mu}\phi_{b}{)}(\partial^{\mu}\phi_b)\right]
+{\mu_I^2\over6f^2}\phi_a\phi_a\left[\phi_2^2+\phi_3^2\right]\;.
\eqa
The NLO Lagrangian Eq.~(\ref{l4}) is expanded to second order in $\phi_a$, which gives
\bqa
{\cal L}^{\rm (2)}_4&=&-2(l_1+l_2){\mu_I^4\over f^2}(\phi_2^2+\phi_3^2)
+2(2l_1+l_2){\mu_I^2\over f^2}(\partial_0\phi_1)^2
+2l_1{\mu_I^2\over f^2}(\partial_{\mu}\phi_a)(\partial^{\mu}\phi_{{a}}) 
+2l_2{\mu_I^2\over f^2}\left[
(\partial_{\mu}\phi_1)(\partial^{\mu}\phi_1)+(\partial_{0}\phi_a)^2
\right]\;.
\eqa
The order-$p^6$ contributions to the pressure from one-loop graphs with counterterm insertions and two-loop graphs are
\bqa
{\cal P}_2^{\rm loops}&=&
{\mu^2_I\over6f^2}\left[3I^2_1(m^2_2)+3I^2_1(m^2_3)+2I_1(m^2_2)I_1(m^2_3)\right]
-{1\over6f^2}\left[m^2_2+m^2_3\right]I_1(m^2_2)I_1(m^2_3)
\nonumber\\&&{-}{\mu^2_I\over f^2}\left[J(m^2_2)+J(m^2_3)\right]
-{2}(l_1+l_2){\mu^4_I\over f^2}\left[I_1(m^2_2)+I_1(m^2_3)\right]
\nonumber\\&&
+{2}l_1{\mu^2_I\over f^2}\left[m^2_2 I_1(m^2_{2})+m^2_3 I_1(m^2_3)\right]
+{2}l_2{\mu^2_I\over f^2}\left[{m^2_2\over d+1}I_1(m^2_2)+{m^2_3\over d+1}I_1(m^2_3)\right]\;,
\label{omegaeffnnlo}
\eqa
\end{widetext}
where the integral $J(m^2)$ is defined in Eq.~(\ref{j2def}).
Counterterm diagrams with a massless propagator or double-bubble diagrams
with a massless propagator vanish in dimensional regularization since there is
no mass scale in the corresponding integrals. Contributions from these diagrams are not included in Eq.~(\ref{omegaeffnnlo}) above.
Using the fact that $m_2=m_3=\mu_I$ and 
the expression for $J(m^2)$ in Eq.~(\ref{jexp}),
Eq.~(\ref{omegaeffnnlo}) reduces to 
\bqa
\label{nnlo}
{\cal P}_2^{\rm loops}&=&{d-1\over d+1}{\mu_I^2\over f^2}I_1^2(\mu_I^2)
-l_2{4d\over d+1}{\mu_I^4\over f^2}I_1(\mu_I^2)\;.
\eqa
Note that the dependence on $l_1$ drops out.
The contribution to the pressure from the static part of ${\cal L}_6$ is
\bqa
{\cal P}_2^{\rm static}&=&2(C_{24}+C_{25}+C_{26})\mu_I^6\;.
\label{lstat2}
\eqa
Adding Eqs.~(\ref{nnlo}) and~(\ref{lstat2}), renormalizing $\mathcal{C}=C_{24}+C_{25}+C_{26}$
according to Eq.~(\ref{cjdef}), we obtain
the NNLO contribution to the pressure. The coefficient $a_2$ reads
\bqa
\nonumber
a_2&=&2C_r-{1\over(4\pi)^2}\left[{1\over2}-3\log{\Lambda^2\over\mu_I^2}\right]l_2^r
\\ &&
+{1\over(4\pi)^4}\left[{-{1\over24}}
-{1\over3}\log{\Lambda^2\over\mu_I^2}+{1\over2}\log^2{\Lambda^2\over\mu_I^2}
\right]\;.
\label{a2}
\eqa
It can be verified that $a_2$ is independent of the scale $\Lambda$ using
the running of $l_2^r(\Lambda)$ and $C^r(\Lambda)$. Comparing the coefficients $a_1$ and $a_2$, we note that the effective expansion parameter is $\mu_I^2/(4\pi)^2f^2$. This suggests that the chiral limit should be a good approximation for $m_{\pi}\ll\mu_I\ll4\pi f$.

\section{Low-energy effective theory and phonon damping rate}
\label{loweft}
In section~\ref{lores}, we calculated the dispersion relations for the charged and neutral
mesons in the pion-condensed phase at leading order in the low-energy expansion.
The Goldstone mode has a linear dispersion relation for
small momentum $p$, which follows directly from a Taylor expansion of the dispersion relation Eq.~(\ref{E2p}) around $p=0$,
\bqa
\label{phono}
E_{\tilde{\pi}^-}(p)&=&\sqrt{\mu_I^4-m_{\pi,0}^4\over\mu_I^4+3m_{\pi,0}^4}p+{\cal O}(p^2)\;,
\eqa
where $m_{\pi,0}$ is the mass of the charged pion mass at tree level. More generally, one can ask about the low-energy dynamics and the low-energy effective theory that describes the Goldstone boson or phonon alone. Let us for simplicity discuss the two-flavor case.
Since the massive excitations have  masses of $\mu_I$ and $\mu_I\sqrt{\mu_I^4+{3m_{\pi,0}^4}\over\mu_I^4}$ in the broken phase, the low-energy effective theory will be valid for $p\ll \mu_I$. In the case of a single chemical potential $\mu$ and a Goldstone boson $\phi$
associated with the breaking of a $U(1)$ symmetry, Son showed how to construct such a theory more than two decades ago~\cite{lowson}. The prescription is remarkably simple: 
The effective theory for the GB $\phi$ is given in terms of the thermodynamic pressure ${\cal P}$ as a function of the chemical potential {$\mu$}
and possibly other quantities such as meson masses simply by making the substitution $\mu\rightarrow\sqrt{\nabla_{\mu}\phi\nabla^{\mu}\phi}$, i.e. 
\bqa
\label{substi}
{\cal L}&=&{\cal P}(\mu\rightarrow\sqrt{\nabla_{\mu}\phi\nabla^{\mu}\phi})\;,
\eqa
where the covariant derivative is $\nabla_{\mu}\phi=\partial_{\mu}\phi-\delta_{\mu0}\mu$.
The only assumption that was made is that the dispersion relation for the phonon
is linear. Eq.~(\ref{phono}) is only linear for small momenta and once there are sizable corrections, the effective theory breaks down.~\footnote{For a color superconductor, the momenta must be much smaller than the superconducting gap $\Delta$. For a dilute Bose see section~\ref{dilute}.} Making the substitution {Eq.~(\ref{substi})}
in the LO pressure Eq.~(\ref{pppion}) with $\Delta m^2_{\rm EM}=0$,
expanding the Lagrangian in powers of derivatives, and rescaling the field, we obtain 
\bqa
\nonumber
{\cal L}&=&{1\over2}(\partial_0\phi)^2-{1\over2}c_s^2(\nabla\phi)^2+c_1(\partial_0\phi)^3
+c_1\partial_0\phi(\nabla\phi)^2+\cdots\;,
\\
\label{expansionl2}
\eqa
where we have omitted a linear term and the ellipses indicate higher-order operators.
The phonon speed $c_s$ and the coupling $c_1$ are
\bqa
\nonumber
c_s=\sqrt{\mu_I^4-m_{\pi,0}^4\over\mu_I^4+3m_{\pi,0}^4}\;,
c_1={2m_{\pi,0}^4\mu_I\over f}{1\over(\mu_I^4+3m_{\pi,0}^4)^{3\over2}}\;.
\\ &&
\label{cs}
\eqa
We next consider loop corrections to the dispersion relation. 
In order to calculate the full self-energy at one loop, we need the quartic terms
in the expansion Eq.~(\ref{expansionl2}), however, the corresponding diagrams contribute only to its real part. The corresponding correction is simply a correction to the phonon speed which can also be calculated directly from the equation of state. In order to calculate the imaginary part and hence the damping rate, it is sufficient to consider the cubic terms. Our calculations closely follow those of Refs.~\cite{cfl,cfl2,nlonr}, in which the leading-order phonon speed and the damping rate in the color-flavor-locked phase of QCD were calculated. The expression for the relevant self-energy diagram is given by the following integral in Euclidean space
\bqa
\Pi(P)&=&c_1^2\int_Q{F(P,Q)\over(q_0^2+c_s^2q^2)[(p_0-q_0)^2+c_s^2({\bf p}-{\bf q})^2]}\;,
\eqa
where the function $F(P,Q)$ is defined as
\bqa
\nonumber
F(P,Q)&=&2\left\{3(p_0-q_0)p_0q_0-(p_0-q_0){\bf p}\cdot{\bf q}
\right. \\ && \left.
{-p_0}({\bf p}\cdot{\bf q}-q^2)
{-q_0}(p^2-{\bf p}\cdot{\bf q})
\right\}^2\;.
\eqa
The integral is next rewritten using Feynman parameters, changing the variables
$R=Q-Px$ and scaling $r_0$, $r_0\rightarrow r_0/c_s$.
The leading contribution is obtained by setting $R=0$ in the numerator. 
The function $F(P,R)$ then reduces to $G(P)=F(P,0)$, where
\bqa
G(P)&=&18x^2(1-x)^2p_0^2(p_0^2+p^2)^2\;.
\eqa
This yields
\bqa
\Pi(P)&=&{c_1^2\over c_s^3}\int_0^1dx\int_R{G(P)\over\left[R^2+{p_0^2+c_s^2p^2\over c_s^2}x(1-x)\right]^2}\;,
\eqa
We next use Eq.~(\ref{i222}) for the integral of momenta $R$.
Integrating the resulting expression with respect to $x$ and going back to Minkowski space, ${ip_0\rightarrow\omega+i\epsilon}$ yields
\bqa
\nonumber
\Pi(\omega,p)&=&-{3c_1^2\over5(4\pi)^2c_s^3}\omega^2(\omega^2-p^2)^2
\\ &&\times
\left[
{1\over\epsilon}-
\log{\omega^2-c_s^2p^2\over c^2_s\Lambda^2}+{47\over30}+i\pi\right]\;.
\eqa
The damping rate $\gamma$ is defined as
\bqa
\gamma&=&-{{\rm Im}\Pi({\omega,p})\over\omega}\bigg|_{\omega^2=c_s^2p^2}\;.
\eqa
In the nonrelativistic limit, 
$c_s^2\ll1$, and we can approximate $(\omega^2-p^2)^2$ by $p^4$, which yields
\bqa
\gamma&=&{3c_1^2\over160\pi c_s^2}p^5\;.
\eqa
In the nonrelativistic limit, $c_s^2\rightarrow{n_I\over4m_{\pi^{},0}f^2}$ and $c_1\rightarrow{1\over4fm_{\pi^{},0}}$, which follows from Eq.~(\ref{cs}) and the results on the dilute Bose gas in the next section. This yields
\bqa
\gamma&=&{3p^5\over640\pi m_{\pi^{},0}n_I}\;,
\eqa
which is the classic result by Beliaev~\cite{beli} for a dilute Bose gas. In the next section, we discuss the dilute Bose gas and the nonrelativistic limit of $\chi$PT further.

\section{Dilute Bose gas and the nonrelativistic limit of $\chi$PT} 
\label{dilutie}
In this section, we first briefly discuss the classic textbook example of Bose condensation, namely that of a nonrelativistic dilute Bose gas~\cite{fetter}. The leading correction to the energy density is derived using effective field theory methods. We then show that a pion condensate behaves nonrelativistically close to the phase transition~\cite{heman}. Finally, we recover the two-flavor results from Ref.~\cite{zhou} by taking the limit $m_s\rightarrow\infty$.

\subsection{Dilute Bose gas}
\label{dilute}

The dilute Bose gas has been studied extensively for several decades beginning with the paper by Bogoliubov~\cite{bogo} in 1947. The starting point is a nonrelativistic low-energy effective field theory that describes the particles at momenta much lower than the inverse 
of the range of their interactions~\cite{braaten}. The Lagrangian that describes the system at finite number density is
\bqa
\nonumber
{\cal L}&=&\psi^{\dagger}(i\partial_0+\mu_{\rm NR})\psi
-{1\over2m}\nabla\psi^{\dagger}\cdot\nabla\psi-{1\over4}g
(\psi^{\dagger}\psi)^2
\\ &&
-{1\over4}h[\nabla(\psi^{\dagger}\psi)]^2-{1\over36}g_3(\psi^{\dagger}\psi)^3+\cdots\;,
\label{nrlag}
\eqa
where the quantum field $\psi^{\dagger}$ creates a particle, $\psi$ destroys a particle, $\mu_{\rm NR}$ is the nonrelativistic chemical potential, $g$, and $g_3$ are coupling constants. 
The ellipses indicate terms that are of higher order in the number fields $\psi$, $\psi^{\dagger}$ and/or their derivatives.
The term $(\psi^{\dagger}\psi)^2$ represents $2\rightarrow2$ scattering and the
coupling $g$ is related to the $s$-wave scattering length $a$ as $g={8\pi a\over m}$.
The term ${1\over4}h[\nabla(\psi^{\dagger}\psi)]^2$
includes the effective range $r_s$ of the two-body potential, where 
$h=2\pi a^2r_s$. The term $(\psi^{\dagger}\psi)^3$ represents $3\rightarrow3$ scattering.

At zero temperature, the expansion parameter of the dilute Bose gas is the so-called 
(dimensionless) gas parameter $\sqrt{na^3}$, where $n$ is the number density.
Bogoliubov~\cite{bogo} obtained the mean-field results for the pressure, number density,  and energy density. For example, the energy density is ${\cal E}(n)={2\pi a n^2\over m}$.
The leading corrections to Bogoliubov's result for the energy density
were calculated by Lee, Huang, and Yang~\cite{LN,LN0} for a hard-sphere potential.
Later, part of the next-to-leading order correction was calculated 
by Wu~\cite{wu}, by Hugenholz and Pines~\cite{hug}, and by Sawada~\cite{saw}. 
A complete next-to-next-to-leading order result was obtained by Braaten and Nieto~\cite{braaten}
using effective-field theory methods. The result depends not only on the scattering
length $a$, but also on an energy-independent term in the
scattering amplitude for $3\rightarrow3$ scattering. The result is
\bqa
{\cal E}(n)&=&{2\pi a n^2\over m}\left[1+{128\over15\sqrt{\pi}}\sqrt{na^3}
\right.\nonumber\\&&\left.
+\left({32\pi-24\sqrt{3}\over3}\log{na^3}+\mathcal{G}\right)na^3\right]\;,
\label{bosongas}
\eqa
where $\mathcal{G}$ is a constant involving the coupling $g_3$. It was already realized by Hugenholz and Pines that physical quantities depend on 
other quantities than the $s$-wave scattering length $a$.
These effects are referred to as nonuniversal effects and are mimicked by e.g. 
the terms ${1\over4}h[\nabla(\psi^{\dagger}\psi)]^2$ and
$g_3(\psi^{\dagger}\psi)^3$ in Eq.~(\ref{nrlag}).
A detailed discussion of nonuniversal effects involving these terms and a
comparison with diffusion Monte Carlo calculations~\cite{latbec} can be found in Ref.~\cite{hermans}.

We now rederive the first two terms in the expansion Eq.~(\ref{bosongas}) using the
effective nonrelativistic Lagrangian Eq.~(\ref{nrlag}).
The first term is the mean-field result, while the second arises from a one-loop calculation.
The complex field is written as $\psi=v+\tilde{\psi}$, where
$v=\langle\psi\rangle$ is its expectation value and $\tilde{\psi}$ is a fluctuating 
quantum field with $\langle\tilde{\psi}\rangle=0$. The fluctuating field is written as
$\tilde{\psi}={1\over\sqrt{2}}(\psi_1+i\psi_2)$, where $\psi_1$ and $\psi_2$ are real fields. 
To second order in the fluctuations, one finds
\bqa
{\cal L}^{(0)}&=&\mu_{\rm NR}v^2-{1\over4}gv^4\;,
\\
{\cal L}^{\rm (1)}&=&{vX\over\sqrt{2}m}\psi_1\;,\\
{\cal L}^{\rm (2)}&=&{1\over2}(\dot{\psi}_1\psi_2-\psi_1\dot{\psi}_2)
+{1\over4m}\psi_1\left(\nabla^2+Y\right)\psi_1
\nonumber
\\ &&
+{1\over4m}\psi_2\left(\nabla^2+X\right)\psi_2\;,
\eqa
where the superscript indicates the power of the field, a dot means a time derivative, $X=2m(\mu_{\rm NR}-{1\over2}gv^2)$, and $Y=2m(\mu_{\rm NR}-{3\over2}gv^2)$. The propagator matrix is 
\bqa
\nonumber
D(\omega,p)={i\over\omega^2-E^2(p)+i\epsilon}
\begin{pmatrix}
{1\over2m}\left(p^{2}-X\right)&-i\omega\\
i\omega&{1\over2m}\left(p^{2}-Y\right)\\
\end{pmatrix}\;,
\\
&&
\eqa
where the spectrum is
\bqa
E(p)&=&{1\over2m}\sqrt{(p^2-X)(p^2-Y)}\;.
\label{spectri}
\eqa
The thermodynamic potential in the mean-field approximation is as usual given by minus the static part of the Lagrangian,
\bqa
\Omega_0(\mu_{\rm NR},v)&=&-\mu_{\rm NR}v^2+{1\over4}gv^4\;.
\eqa
The minimum of $\Omega_0(\mu_{\rm NR},v)$ is $v_0=\sqrt{{2\mu_{\rm NR}\over g}}$. At the minimum $v_0$, $X=0$ and Eq.~(\ref{spectri}) reduces to the Bogoliubov spectrum
$E_p={p\over2m}\sqrt{p^2+4m\mu_{\rm NR}}$. The dispersion relation is linear for small momenta, $p^2\ll4m\mu_{\rm NR}$, and that of a free nonrelativistic particle for large momenta, 
$p^2\gg4m\mu_{\rm NR}$.

In order to calculate the NLO corrections to the thermodynamic quantities, we need a few divergent loop integrals. We use dimensional regularization to regulate these integrals. The integrals needed were introduced in Ref.~\cite{effective} and are of the form
\bqa
I_{m,n}(M^2)&=&\Lambda^{2\epsilon}\int{d^dp\over(2\pi)^d}
{p^{2m}\over p^n\left(p^2+M^2\right)^{{n\over2}}}\;,
\label{defigam}
\eqa
They satisfy the recursion relation
\bqa
\label{rec}
{dI_{m,n}{(M^2)}\over dM^2}&=&-{1\over2}nI_{m+1,n+2}(M^2)\;,
\eqa
which follows directly from the definition Eq.~(\ref{defigam}).
Evaluating the integrals in dimensional regularization, we find
\bqa
\nonumber
I_{m,n}(M^2)&=&{M^{3+2m-2n}\over(4\pi)^{{{d}\over2}}}
\left({\Lambda\over M}\right)^{2\epsilon}
\\ &&\times
{\Gamma({d-n\over2}+m)\Gamma(n-m-{d\over2})\over\Gamma({n\over2})\Gamma({d\over2})}\;.
\eqa
We specifically need
\bqa
\label{01}
I_{0,-1}(M^2)&=&{16\over15}{M^5\over(4\pi)^2}
\left[1+{\cal O}(\epsilon)\right]
\;,\\
\label{11}
I_{1,1}(M^2)&=&{16M^3\over3(4\pi)^2}\left[1+{\cal O}(\epsilon)\right]\;.
\eqa
The integrals are finite in the limit $d\rightarrow3$ reflecting that the 
ultraviolet divergences are powerlike.

The NLO pressure is given by the NLO thermodynamic potential evaluated at the classical minimum $v_0$ cf. Eq.~(\ref{ponshell}). This is convenient since $X=0$. The NLO pressure then becomes
\bqa
\nonumber
{\cal P}(\mu_{\rm NR})&=&-\Omega_{0}(\mu_{\rm NR},v_0)-\Omega_1(\mu_{\rm NR},v_0)
\\ \nonumber
&=&{\mu_{\rm NR}^2\over g}-{1\over2}\int_pE(p)
\\ \nonumber
&=&{\mu_{\rm NR}^2\over g}-{1\over4m}I_{0,-1}(M^2)\\
&=&
{\mu_{\rm NR}^2\over g}\left[1-{16(4m)^{{3\over2}}\sqrt{\mu_{\rm NR}g^2}\over15(4\pi)^2}\right]\;,
\label{ppnr}
\eqa
where 
$M^2=4m\mu_{\rm NR}$, and we have used Eq.~(\ref{01}).
The number density is then given by
\bqa
n(\mu_{\rm NR})&=&
{2\mu_{\rm NR}\over g}-{1\over2}I_{1,1}(M^2)\;,
\label{nnr}
\eqa
where we have used the recursion relation Eq.~(\ref{rec}).
We can invert Eq.~(\ref{nnr}) to obtain the chemical potential in terms of the number density. To the order we are calculating, we can make the substitution
$\mu_{\rm NR}\rightarrow{1\over2}gn$ in the loop integral $I_{1,1}(M^2)$. This yields
\bqa
\nonumber
\mu_{\rm NR}(n)&=&{1\over2}gn+{1\over4}gI_{1,1}(2mgn)
\\
&=&{4\pi an\over m}\left[1+{32\over{3}\sqrt{\pi}}\sqrt{na^3}\right]\;,
\eqa
where we have used Eq.~(\ref{11}) and $g={8\pi a\over m}$ in the last line.
The energy density is then
\bqa
\nonumber
{\cal E}(n)&=&-{\cal P}+\mu_{\rm NR}n
\\
&=&{1\over4}gn^2+{1\over4m}I_{0,-1}(2mgn)\;.
\eqa
Note that, to the order we are calculating, the terms involving $I_{1,1}(\mu_{\rm NR})$ cancel
in final result for the energy density. Using the result Eq.~(\ref{01}) for the integral and the 
expression for $g$ in terms of the $s$-wave scattering length, we obtain the result
of Lee, Huang and Yang~\cite{LN,LN0},
\bqa
{\cal E}(n)&=&{2\pi a n^2\over m}\left[1+{128\over15\sqrt{\pi}}\sqrt{na^3}\right]\;.
\label{bosongas2}
\eqa

\subsection{Nonrelativisit limit of $\chi$PT}
In this section, we take the nonrelativistic limit of chiral perturbation theory in order to make contact with the theory of dilute Bose gases discussed in the previous section. In order to do so, we introduce the nonrelativistic chemical potential $\mu_{\rm NR}$ by writing $\mu_I=m_{\pi}+\mu_{\rm NR}$, where $m_{\pi}$ is the physical pion mass. In a consistent calculation, $m_{\pi}$ must be calculated
to the same order in the low-energy expansion as the pressure itself.
Expanding the pressure Eq.~(\ref{nlop}) to order $\mu_{\rm NR}^{5/2}$, we obtain
\begin{widetext}
\begin{align}
\nonumber
{\cal P}
&=2f^2\mu^2_{\rm NR}\left\{1
-{3\over2}{\delta m_{\pi}^2\over m_{\pi,0}^2}
+\left[32L_1^r+32L_2^r+16L_3^r-40L_4^r-20L_5^r+80L_6^r+40L_8^r\right.\right.
\\\nonumber & \left.\left. 
+{1\over(4\pi)^2}\left({11\over4}\log{\Lambda^2\over m_{\pi,0}^2}
+{1\over2}\log{\Lambda^2\over m_{K,0}^2}
+{5\over36}\log{\Lambda^2\over m_{\eta,0}^2}+{4\over9}
\right)\right]{m_{\pi,0}^2\over f^2}
\right. \\ &\left.
+\left[-8L^r_4+48L^r_6+{1\over(4\pi)^2}\left(
\log{\Lambda^2\over m_{K,0}^2}+{1\over3}\log{\Lambda^2\over m_{\eta,0}^2}\right)\right]{\tilde{m}_{K,0}^2\over f^2}
\right\}
-{64m_{\pi,0}\mu_{\rm NR}^2\sqrt{4m_{\pi,0}\mu_{\rm NR}}\over15(4\pi)^2}
\;,
\label{p3f0}
\end{align}
where the term $-{3\over2}{\delta m_{\pi}^2\over m_{\pi,0}^2}$ arises 
from distinguishing between $m_{\pi}$ and $m_{\pi,0}$ in the tree-level contribution to the pressure (which is necessary for a consistent calculation). This term can be read off 
Eq.~(\ref{mpi}). There is also a term linear in $\mu_{\rm NR}$ from the tree-level contribution to the pressure for the same reason. This term is cancelled by a similar term from loop corrections. Thus the first term in the expansion is quadratic in $\mu_{\rm NR}^2$.
Also note the last term in Eq.~(\ref{p3f0}), which comes from the hypergeometric function. This is exactly the loop correction in Eq.~(\ref{ppnr}). This yields
\bqa
\nonumber
{\cal P}
&=&2f^2\mu^2_{\rm NR}\left\{1
+\left[32L_1^r+32L_2^r+16L_3^r-16L_4^r-8L_5^r+32L_6^r+16L_8^r\right.\right.
\\\nonumber && \left.\left. 
+{1\over(4\pi)^2}\left({7\over2}\log{\Lambda^2\over m_{\pi,0}^2}
+{1\over2}\log{\Lambda^2\over m_{K,0}^2}
+{1\over18}\log{\Lambda^2\over m_{\eta,0}^2}+{4\over9}
\right)\right]{m_{\pi,0}^2\over f^2}
+\left[16L^r_4+{1\over(4\pi)^2}
\log{\Lambda^2\over m_{K,0}^2}\right]{\tilde{m}_{K,0}^2\over f^2}
\right\}
\\ &&
-{64m_{\pi,0}\mu_{\rm NR}^2\sqrt{4m_{\pi,0}\mu_{\rm NR}}\over15(4\pi)^2}\;,
\label{p3f}
\eqa
where we have defined $\tilde{m}_{K,0}=B_0m_s$, i.e. the bare kaon mass in the limit 
of large $m_s$. This form is particularly convenient if we are interested in this limit.
Expanding Eq.~(\ref{p3f}) in powers of $1/m_s$, using Eqs.~(\ref{tildef})--(\ref{l3def}),
we obtain
\bqa
{\cal P}&=&2\tilde{f^2}\mu_{\rm NR}^2\left\{1
+\left[
8l_1^r+8l_2^r+2l_3^r+{1\over(4\pi)^2}\left({7\over2}\log{\Lambda^2\over m_{\pi,0}^2}+
{1\over2}\right)\right]{m_{\pi,0}^2\over f^2}
\right\}
-{64m_{\pi,0}\mu_{\rm NR}^2\sqrt{4m_{\pi,0}\mu_{\rm NR}}\over15(4\pi)^2}\;.
\eqa
\end{widetext}
Using the definition Eq.~(\ref{lr2}) replacing the running parameters $l_i^r$ by their counterparts $\bar{l}_i$ yields
\bqa
{\cal P}&=&{m_{\pi}\over8\pi a}\mu_{\rm NR}^2\left[1
-{32\over15\pi}\sqrt{4m_{\pi}\mu_{\rm NR}a^2}
\right]\;,
\eqa
where we have used the two-flavor expression for the pion mass to one-loop order and the scattering length $a=-a_0^2/m_{\pi}$, where~\cite{gasser2}
\bqa
\label{nlompi}
m_{\pi}^2&=&m^2_{\pi,0}\left[1-{m^2_{\pi,0}\over2(4\pi)^2f^2}\bar{l}_3\right]\;,\\
a_0^2&=&-{m^2_{\pi,0}\over4(4\pi)f^2}\left[
1-{4m^2_{\pi,0}\over3(4\pi)^2f^2}\left(\bar{l}_1+2\bar{l}_2+{3\over8}\right)\right]\;.
\label{a00}
\eqa
Calculating the isospin density and the energy density, we find
\bqa
n_I&=&{m_{\pi}\over4\pi a}\mu_{\rm NR}\left[1
-{8\over3\pi}\sqrt{4m_{\pi}\mu_{\rm NR}a^2}\right]
\;,\\
{\cal E}(n_I)&=&m_{\pi}n_I+
{2\pi a n_I^2\over m}\left[1+{128\over15\sqrt{\pi}}\sqrt{n_Ia^3}\right]\;.
\label{energynr}
\eqa
We can now compare our result Eq.~(\ref{energynr}) with the
result of Braaten and Nieto, Eq.~(\ref{bosongas}).
The first term in Eq.~(\ref{energynr}) is the contribution to ${\cal E}$ associated with the rest mass $m_{\pi}$ of the boson.
This term is absent in Eq.~(\ref{bosongas}) since it is automatically removed by subtracting the rest mass energy in the nonrelativistic Lagrangian Eq.~(\ref{nrlag}). Omitting this term, Eq.~(\ref{energynr}) is the same as the  first and second term in Eq.~(\ref{bosongas}), i.e. 
there is agreement to one-loop order.
The last term in Eq.~(\ref{bosongas}) involves a two-loop calculation and is therefore not included in Eq.~(\ref{energynr}). See also Ref.~\cite{newnic} for similar results.

\section{Summary and Outlook}
\label{summary}
In this paper, we have discussed various aspects of Bose condensation in QCD
at finite $\mu_I$ and $\mu_S$
using chiral perturbation theory, which is the low-energy effective theory describing the pseudo-Goldstone bosons.
We have been focusing on the pion-condensed phase mainly due to the fact
that in this case it is possible to compare our predictions with those of lattice QCD.
However, with relatively little effort similar results for the kaon-condensed phases
can be obtained. Lattice QCD and $\chi$PT agree very well in the region where the latter is
expected to be valid. Depending on taste, one can view this as a check of $\chi$PT as an effective theory of QCD or a check of the simulations.
Bose-condensation in QCD is a very rich system: in the region $\mu_I\simeq m_{\pi}$, we have made
contact with the dilute Bose gas and the classic results by Bogoliubov, Beliaev, Lee, Yang and Huang, and others. In the ultrarelativistic limit, the Goldstone mode is exactly linear and propagates with the speed of light. 
The speed of sound, $c_s=\sqrt{{\mu_I^4-m_{\pi,0}^4}\over\mu_I^4+3m_{\pi,0}^4}$,
is a measure of how relativistic the system is.

The present work can be extended in several directions. Firstly, it would be interesting to calculate the thermodynamic quantities and the phase diagram to order ${\cal O}(p^4)$ including electromagnetic interactions. This would require using Urech's next-to-leading order Lagrangian~\cite{urech1} and the evaluation of a very complicated functional determinant.
In the same vein, one could calculate the meson and gauge boson masses to ${\cal O}(p^4)$.
A more straightforward extension would be to finite temperature. In a two-flavor 
calculation~\cite{mojahed}, the critical line between the normal phase and the BEC phase was mapped out in the $\mu_I$-$T$ plane. Good agreement between $\chi$PT and lattice simulations was only found for temperatures up to approximately 30 MeV. Whether the inclusion of heavier mesons would improve the situation is an
open question. 

We have also noticed the disagreement between $\chi$PT and the lattice regarding the 
speed of sound for large values of $\mu_I$, which is caused by the fact that 
$\chi$PT has mesonic bound states as degrees of freedom and not quarks.
This problem was addressed in Ref.~\cite{kojo}, where the two-flavor quark-meson model was investigated at finite isospin and vanishing temperature.
The model is in qualitative agreement with lattice data from the BEC to the BCS regime, suggesting that it captures the correct degrees of freedom.

\section*{Acknowledgements}
Q. Yu and H. Zhou have been supported by the Natural Science Foundation of China under Grant
No.12305091, the Natural Science Foundation of Sichuan Province under Grant No.2024NSFSC1367, and the Research Fund for the Doctoral Program of the Southwest University of Science and Technology under Contract No.23zx7122 and No.24zx7117. J. O. Andersen would like to thank the Niels Bohr
International Academy for kind hospitality during his stay where large part of this work
was carried out.
J. O. Andersen would also like to thank Prabal Adhikari and
Martin Mojahed for earlier collaboration as well as Alberto Nicolis, Alessandro Podo, and
Luca Santoni for useful discussions.
The authors thank Bastian Brandt and 
Gergely  Endr\H{o}di for sharing their old and updated lattice data and for discussions 



\end{document}